\documentclass[11pt,a4paper]{article}
\usepackage{jcappub}

\pdfoutput=1

\newcommand{\MeV}{{\rm MeV}}

\newcommand{\GeV}{{\rm GeV}}
\newcommand{\Tx}{T_\times}
\newcommand{\Mp}{M_{\rm P}}
\def\d{{\rm d}}

\usepackage{ulem}

\usepackage{color}
\definecolor{new}{rgb}{0.0,0.6,0.0}

\title{CMB observables and reheat temperature as a window
to models\\ of inflation and freeze-in dark matter production}

\author{Anish Ghoshal,}

\emailAdd{anish.ghoshal@fuw.edu.pl}

\author{Pawe\l\ Koz\'ow,}
\emailAdd{pawel.kozow@fuw.edu.pl}

\author{Marek Olechowski,}
\emailAdd{marek.olechowski@fuw.edu.pl}

\author{Stefan Pokorski}
\emailAdd{stefan.pokorski@fuw.edu.pl}

\affiliation{Institute of Theoretical Physics, Faculty of Physics, University of Warsaw,\\  ul.~Pasteura 5, 02-093 Warsaw, Poland}

\abstract{
A systematic approach is presented for using CMB  observables and reheating temperature for  discriminating between various models of inflation and certain freeze-in dark matter scenarios. It is applied to several classes of $\alpha$-attractor  models as an illustrative example. In the first step, all independent parameters of the inflationary potential are expressed in terms of the  CMB observables (the three parameters - by the scalar spectral index $n_s$, scalar amplitude $A_s$ and the tensor-to-scalar amplitude ratio $r$). For a standard reheating mechanism characterized by the inflaton equation of state  parameter $w$ and its effective dissipation rate $\Gamma$ the reheating temperature is uniquely fixed in terms of the CMB observables  measured for some pivot scale $k_*$. There are striking consequences of this fact. The model independent bounds on the reheating temperature, the BBN lower bound and the upper bound of the order of the GUT/Planck scale, translate themselves for each class of models into very narrow  ranges of the allowed values  of the spectral index $n_s(k_*)$,  providing their strong tests by the present and future CMB data. 
The recent tension between Planck and DESI-ACT results has strong impact on our conclusions.
Furthermore, given a class of inflaton models satisfying those tests, the reheating temperature is an interesting portal to link  the CMB observables to the particle physics scenarios that are sensitive to it. As an example, non-thermal dark matter (DM) production mechanisms are discussed. One obtains then a consistency check between theories of inflation and DM production. If the future precision of the CMB data will constrain the reheating temperature beyond the model independent bounds, further constraints on the DM production will follow. 
}

\begin{document}

\maketitle

\section{Introduction}
The standard paradigm is that, in its early evolution, the universe underwent several distinct phases. Its accelerated expansion (inflation) was followed by a period when the energy stored in the inflaton field was converted to relativistic particles, broadly referred to as a reheating period, during the oscillations of the inflaton field around the minimum of its potential.
After that, the standard radiation-dominated (and later, matter-dominated) evolution of the universe took place. 

It has been emphasized for a long time that the CMB data give us important insight into the inflationary and reheating periods, constraining theoretical models of both \cite{Mukhanov:2005sc}.  
In this paper we 
address similar issues but with a strong emphasis on the reheating temperature  as a crucial parameter for
a systematic approach in using CMB data for  testing and discriminating between different models of inflation.  We show that already the model independent bounds on the reheating temperature, the BBN lower bound, arising due to the precision measurements of primordial abundance of light nuclei, and the upper bound of the order of the GUT/Planck scale \cite{Weinberg:2008zzc}, 
translate themselves into strong constraints on the CMB observables in a given model of inflation. Comparing them with  the present and future CMB experimental data has a strong discriminative power.  Secondly,   the  reheating temperature is  a portal  to constraining particle physics scenarios that are sensitive to it.  As an example we discuss  freeze-in DM scenarios. 
Therefore,  with increased precision of the CMB data,  once stronger than the model independent bounds on the reheating temperature are obtained,  in a given  inflationary model such particle physics scenarios will be tested by the CMB data.  We propose a systematic procedure to determine the reheating temperature, for a given  model of inflation,  directly in terms of the CMB observables.

As the first step, all  independent parameters of an inflaton potential are expressed in terms of the CMB observables such as: the scalar metric perturbation amplitude $A_s(k)$, the scalar spectral index $n_s(k)$, its running 
$\alpha_s(k)$ and the ratio of the tensor to scalar perturbation amplitudes $r(k)$ etc. Here $k=k_*$ is a comoving wavenumber of the metric perturbation, measured by CMB experiments at some physical scale $\lambda_{\rm phys}=a_0/k_*$  determined by the experimental angular resolution (the Planck pivot scale is $k_* = a_0 \times 0.05$ Mpc$^{-1}$) where $a_0$ denotes the Friedman-Laimetre-Robertson-Walker (FLRW) scale factor today. The number of observables has of course to match the number of independent parameters of the potential and their best choice depends on the precision of their experimental values. As an illustrative example we apply our approach to several classes of $\alpha$-attractor models, the so-called E, T and P-models, having three independent parameters, and choosing $n_s, A_s, r$ as the three input CMB observables.
Those potentials are  well motivated in several cosmological scenarios. Some well known cases are the famous Starobinsky inflation  \cite{Starobinsky:1980te}, the Goncharov-Linde model  \cite{Goncharov:1984jlb,Linde:2014hfa}, the Higgs Inflation scenario  \cite{Bezrukov:2007ep}, several superstring-inspired scenarios, \cite{Cicoli:2008gp,Ferrara:2016fwe,Kallosh:2017ced,Kallosh:2017wku})  and no-scale supergravity ~\cite{Ellis:2019bmm,Ellis:2019hps}.

\begin{figure}[t]
    \begin{center}
        \begin{minipage}{0.49\textwidth}
            \includegraphics[width=1.0\textwidth]{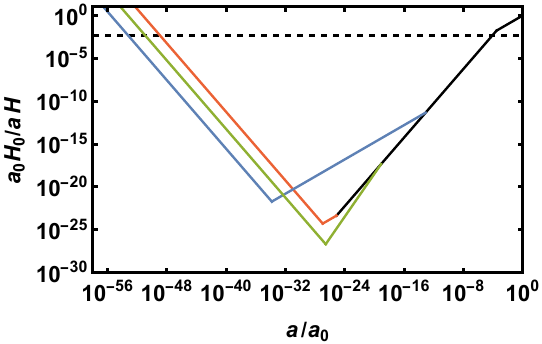}
        \end{minipage}
        
    \end{center}
    \caption{\label{fig:aHinvewol}A schematic evolution of comoving Hubble radius $aH$ as a function of the scale factor $a$. The blue line is for $r(k_*)=0.04$ and $w =0$; the green line is for
    $r(k_*)=10^{-6}$ and $w =1/2$; the  red line is for $r(k_*)=10^{-10}$ and $w =0$; the obtained reheating temperatures are 10 GeV, $10^6$ GeV and $10^{12}$ GeV, respectively. The horizontal dotted line corresponds to the comoving wavenumber value $k_*=a_0\times 0.05\,$Mpc$^{-1}$.
    }
\end{figure}

In the next step we adopt a model for the reheating period. In the standard approach it is described by the inflaton equation of state parameter $w$  and a single new parameter, the inflaton effective dissipation rate $\Gamma$, or equivalently,  by the reheating temperature at the beginning of the radiation era. It turns out that the value of $\Gamma$, or equivalently and more importantly the reheating temperature, are uniquely fixed in terms
of the chosen (three for the $\alpha$-attractor models) CMB observables. 
The logic of this procedure is nicely illustrated by Figure~\ref{fig:aHinvewol}, which  shows a sketch of the evolution of the comoving Hubble scale $(H^{-1})_{com}= (aH)^{-1}$ as a function of the expansion of the universe described by the FLRW scale factor $a(t)$ (quantities on both axes are normalized by their present values).
During inflation, the Hubble scale $H^{-1}$ approximately corresponds to the size of causally connected universe (event horizon) and during the decelerated expansion it approximately gives the size of the particle horizon.
The perturbation with the comoving wavenumber $k_*$
leaves the causal region during inflation when its physical length $\lambda_{\rm phys}^{\rm inf}=a_{k_*}/k_*=H^{-1}_{k_*}$, hence for the value of the comoving Hubble scale $\frac{1}{H_{k_*}a_{k_*}}=\frac{1}{k_*}$.  
In units of $a_0H_0$ and with $k_*=a_0\times 0.05$ Mpc$^{-1}$ one obtains the horizontal dashed line in Figure~\ref{fig:aHinvewol}. For a chosen inflaton potential, the value of $H_{k_*}$, in our approach, has already been calculated in terms of the CMB observables $A_s$ and $r$. Knowing $H_{k_*}$, one can calculate the value of the scale factor when the perturbation left the horizon, $a_{k_*}/{a_0}=0.05\,$Mpc$^{-1}/H_{k_*}$, which determines the crossing of the line describing inflation in a given model with the dashed line corresponding to $k_*=a_0\times 0.05$\,Mpc$^{-1}$. 
Similarly, the Hubble scale at the end of inflation is expressed for a given model of inflation in terms of the CMB parameters and gives us $a_{\rm end}/a_0$. On the right end of the plot, the running of Hubble scale as a function of $a/a_0$ in the matter domination and radiation domination eras is known. In between, we have a period of reheating which, for a given inflation model, is parameterized by the inflaton equation of state parameter $w$ which determines the slope of a line describing reheating era in Figure~\ref{fig:aHinvewol}.  The reheating temperature (temperature at the end of reheating and the beginning of the radiation domination era), $T_{\rm re}$, is fixed by the consistency condition: the perturbation with the comoving wavenumber $k_*$, that left the horizon at the moment parameterized by the values of the CMB observables, re-entered it at some known time during the radiation domination era.\footnote{Some consequences of this condition for $\alpha$-attractor E- and T-models were investigated in \cite{Mishra:2021wkm,German:2022sjd,Garcia:2023tkk}.
} 
The reheating temperature is therefore determined in terms of the CMB parameters.\footnote{A model in a given class of $\alpha$-attractor potentials has three independent parameters, which can be expressed in terms of the three CMB observables. Our approach can be generalized to models with more parameters, if more CMB observables are used, e.g.~also the running of the spectral index $\alpha_s(k_*)$.}

The model independent bounds on the reheating temperature
give, for each class of models,  very narrow ranges of the spectral index $n_s(k)$ and with interesting dependence on the parameter $r$. That is a strong test of inflationary models by  the present and future CMB data. Moreover, one can investigate the implications for models of inflation  of imposing some constraints on the reheating temperature motivated by various physical arguments. For instance, successful standard thermal leptogenesis ~\cite{Fukugita:1986hr} requires $T_{\rm re}\geq 10^9 $ GeV~\cite{Buchmuller:2004nz}.\footnote{It was shown in ref.~\cite{Ghoshal:2022fud} that in non-thermal leptogenesis during the reheating period, CMB observables could probe the
scales of seesaw and leptogenesis.} 
A different, contradictory, example are upper bounds on reheat temperature  derived when solving the so-called gravitino problem in certain models of supersymmetry breaking in supergravity with $T_{\rm re} \leq 10^{7}$ GeV \cite{Khlopov:1984pf,Moroi:1993mb,Asaka:2000zh,Roszkowski:2004jd,Cerdeno:2005eu,Steffen:2006hw}, and axion and axion-like production as DM during reheating \cite{Elahi:2014fsa,Arias:2022qjt,Ghoshal:2024gai,Cheek:2025gvx}.

Given a class of inflaton models, the reheating temperature links the CMB observables with the physics scenarios sensitive to it. As an example we discuss some non-thermal DM production mechanisms. 
It has long been understood that freeze-in DM production is $T_{\rm re}$-sensitive, provided the process is UV sensitive in the energy range accessible during reheating.
This is the case when the DM production can be discussed in the effective field theory (EFT) approach, with its amplitude following from a dimension higher than four operator suppressed by appropriate powers of the decoupling scale $\Lambda_{\rm DM}> T_{\rm re}$.
This means that a particular choice for the set of parameters associated with dark matter sector, such as the mass of the DM particle and the value of the scale $\Lambda_{\rm DM}$, requires a unique value of reheating temperature of the Universe to yield the correct DM relic abundance.
One obtains then a consistency check between theories of inflation and DM production.
If the future precision of the CMB data will constrain the reheating temperature beyond the model independent bounds, further constraints on the DM production will follow.  
We present our results for a ``prototype'' model,  to demonstrate that potentially useful connection.  Similar analyses can be performed 
for any complete $T_{\rm re}$ sensitive DM production model, with a concrete set of operators and their Wilson coefficients, higher order corrections in EFT etc., properly taken into account.\footnote{While this project was ongoing,  the link between CMB observables and freeze-in DM has been also addressed in ref.~\cite{Biswas:2025adi}.}

The details of our approach and the power of the dependence of the reheating temperature on the CMB observables for discriminating between different models of inflation are discussed in Section~\ref{sec:CMB}.
The role of the reheating temperature $T_{\rm re}$ as a link between inflationary models and some DM production mechanisms is discussed in Section~\ref{sec:DM}. 

\section{CMB, inflaton potentials and reheating temperature}
\label{sec:CMB}

At present, the Planck CMB  data give us the value of the spectral index $n_s(k_*)$
for the scalar perturbation mode with the (pivot) wavenumber $k_*/a_0 =0.05\,$Mpc$^{-1}$, its amplitude  $A_s(k_*)$
\cite{Planck:2018vyg} and the upper bound on the ratio of the tensor to scalar perturbation amplitudes $r(k_*)$. Following the inflation paradigm, the values of those observables depend on the shape of the inflaton potential, its  independent parameters and the value of the inflaton field $\phi_{k_*}$ when the mode $k_*$ exits the horizon during inflation. 

In the slow-roll approximation the observables $n_s(k)$, $r(k)$, and $A_s(k)$, where $k$ is some comoving wavenumber of the metric perturbations, are related to the inflaton potential as follows:
\begin{equation}
\label{eq:ns_r_As}
    \begin{aligned}
        n_s(k) &= 1 - 6\epsilon_k + 2 \eta_k\,,\\
        r(k) &= 16\epsilon_k\,, \\
        A_s(k) &= \frac{V(\phi_k)}{24 \pi^2 \epsilon_k \Mp^4 }\,,
    \end{aligned}
\end{equation}
where $\Mp$ is the reduced Planck mass.
We also include in our discussion the running of the spectral index  defined as
\begin{equation}
\label{eq:alphasdlnk}
    \alpha_s(k) ~=~ \frac{\d \, n_s(k)}{\d \log k} 
    ~=~ \left(16 \epsilon_k\eta_k  - 24\epsilon_k^2 - 2\theta_k^2 \right)\,.
\end{equation}
The parameters $\epsilon_k,\eta_k$ and $\theta_k$ read:
\begin{eqnarray}
\label{eq:slow_roll_params}
        \epsilon_k&=&\left.\frac{1}{2}\Mp^2 \left(\frac{\partial_\phi V(\phi)}{V(\phi)}\right)^2\right|_{\phi=\phi_k}\;, \nonumber\\
        \eta_k &=& \left.\Mp^2\,\frac{\partial^{(2)}_\phi V({\phi})}{V(\phi)}\right|_{\phi=\phi_k}\;,
      \\
    \theta_k^2 &=& \left.\Mp^4\,\frac{\partial_{\phi} V( {\phi})\,\partial_{{\phi}}^{(3)}V({\phi})}{V^2({\phi})}\right|_{\phi=\phi_k}\;.
    \nonumber
\end{eqnarray}

We focus now on three classes of the $\alpha$-attractor inflaton models, the so-called E, T and P-models, with the following respective
potentials \cite{Kallosh:2013lkr,Kallosh:2013hoa,Kallosh:2013yoa,Kallosh:2013pby,Kallosh:2013maa,Galante:2014ifa,Kallosh:2022feu}: 
\begin{equation}
    V\left(\phi,\alpha, n,\Lambda_\mathrm{\rm inf}\right) ~=~ \begin{cases}
        \Lambda_{\mathrm{\rm inf}}^4\left(1-\exp\left[-\sqrt{\frac{2}{3\alpha}}\frac{\phi}{\Mp}\right]\right)^{2n}&\quad\text{E-model}\;,\\        \Lambda_{\mathrm{\rm inf}}^4\left(\tanh\left[ \sqrt{\frac{2}{3\alpha}}\frac{\phi}{\Mp}\right]\right)^{2n}&\quad\text{T-model}\;,  \\ 
         \Lambda_{\mathrm{\rm inf}}^4\frac{\phi^{2n}}{\phi^{2n}+\left( \sqrt{\frac{3\alpha}{2}}\Mp\right)^{2n}}&\quad\text{P-model}\;,
    \end{cases}
    \label{eq:alpha-attractors}
\end{equation}  
where $\Lambda_{\rm inf}$ represents a mass scale that determines the energy scale of the inflation and $\sqrt{\alpha}\Mp$ is an effective scale that can be higher than $\Mp$.\footnote{
We use the same normalization of $\alpha$ in all three models. Observe that it is different from the normalization for T-model used e.g.~in \cite{Kallosh:2013yoa}.
}  
A special case of the E-model, with $\alpha=1$ and $n=1$, mimics the standard Higgs-Starobinsky inflaton potential \cite{Bezrukov:2007ep}. 

In the following we shall present the necessary formulae for the E-model, relegating the analogous ones for the T and P-models to Appendix~\ref{app:TP-models}. For the E-model, all the perturbation parameters $n_s(k)$ etc., expressed in terms of the potential parameters, read (see eqs.~\eqref{eq:ns_r_As}-\eqref{eq:alpha-attractors}):
\begin{eqnarray}
\label{eq:ns_as_fun_of_alfa_n_phik}
n_s(k)&=&n_s(\alpha, n, \phi_k)=1-\frac{8n\left(\exp\left(\sqrt{\frac{2}{3\alpha}}\frac{\phi_k}{\Mp}\right)+n\right)}{3\alpha\left(\exp\left(\sqrt{\frac{2}{3\alpha}}\frac{\phi_k}{\Mp}\right)-1\right)^2}\,,
\\
\label{eq:r_as_fun_of_alfa_n_phik}
  r(k)&=&r(\alpha, n, \phi_k)=\frac{64 n^2}{3\alpha\left(\exp\left(\sqrt{\frac{2}{3\alpha}}\frac{\phi_k}{\Mp}\right)-1\right)^2}\,,
\\
\label{eq:As_as_fun_of_alfa_n_phik_llambdainf}
A_s(k)
&=& A_s(\alpha,n,\phi_k,\Lambda_{\rm inf})=
\nonumber\\
&=&\frac{\alpha}{32\pi^2n^2}\frac{\Lambda^4_{\rm inf}}{\Mp^4}
\exp\left(2\sqrt{\frac{2}{3\alpha}}\,\frac{\phi_k}{\Mp}\right)
\left(1-\exp\left(-\sqrt{\frac{2}{3\alpha}}\,\frac{\phi_k}{\Mp}\right)\right)^{2(n+1)},
\\
\label{eq:alphsS_as_fun_of_alfa_n_phik}
    \alpha_s(k)&=&\alpha_s(\alpha, n, \phi_k)=
       - \frac{32n^2\exp\left(\sqrt{\frac{2}{3\alpha}}\frac{\phi_k}{\Mp}\right)
\left(\exp\left(\sqrt{\frac{2}{3\alpha}}\frac{\phi_k}{\Mp}\right)+2n+1\right)
}{
9\alpha^2\left(\exp\left(\sqrt{\frac{2}{3\alpha}}\frac{\phi_k}{\Mp}\right)-1 \right)^4
}\,.
\end{eqnarray}

The exponent $2n$ in the potential \eqref{eq:alpha-attractors} is an even integer and we consider it as a number which defines the model.\footnote{
We thank Renata Kallosh and Andrei Linde for drawing our attention to the fact that for polynomial potentials like P-model fractional values of $n$ are also of theoretical interest.}
The other parameters $\Lambda_{\rm inf}, \alpha$  and the inflaton field value  $\phi_k$ when the mode with the co-moving wavenumber $k$ left the horizon can be expressed  in terms of the three observables by the inverse relations.
It is convenient to define the following combination of the CMB observables,  $n_s$ and $r$, and the integer potential parameter $n$:
\begin{equation}
\label{eq:xi}
\xi=n\left(8\,\frac{1-n_s}{r}-1\right)\,.
\end{equation}
We get then:
\begin{eqnarray}
\alpha
&=&
\frac{64n^2}{3r}\frac{1}{\left(\xi-1\right)^2}
\,,\\
\label{eq:phi_k}
\phi_k
&=&
\Mp\sqrt{\frac{32}{r}}\,\frac{n}{\left(\xi-1\right)}\ln(\xi)
\,,\\
\Lambda^4_{\rm inf}
&=&
\frac{3\pi^2}{2}\,\Mp^4\,rA_s\left(\frac{\xi}{\xi-1}\right)^{2n}
\label{eq:Lambda_inf}
\,,
\end{eqnarray}
Thus, the values of the three observables for some comoving wavenumber $k$ fully determine the parameters of the $\alpha$-attractor potential for a given choice of the exponent $2n$. In this case the running $\alpha_s(k)$ is the prediction of the model:
\begin{equation}
\alpha_s
=
-\frac{r^2\xi\left(\xi+2n+1\right)}{128n^2}  \,.
\label{eq:alphaS-E}
\end{equation}
In models with four  independent parameters, it would be needed to fix the model parameters.

The observables are measured by the Planck experiment for the comoving wavenumber $k=k_* = a_0\times 0.05$\,Mpc$^{-1}$. The  results are usually presented as the experimentally allowed regions (at some confidence level) in the plane $(n_s(k_*), r(k_*))$
or $(n_s(k_*), \alpha_s (k_*))$. It is customary to compare them with models of inflation by checking if those allowed regions are consistent with the \textbf{assumed} number of e-folds, usually $50\div60$, during the rolling down of the inflaton field from its value $\phi_{k_*}$ to the value $\phi_{\rm end}$ at the end of inflation. The weakness of this method  of comparing the CMB data with the models of inflation is that the a priori acceptable range of $N_k$ can actually be very large and is correlated with the expansion during the reheating period (see e.g.~textbooks \cite{Kolb:1990vq,Weinberg:2008zzc}, it is also evident from Figure~\ref{fig:aHinvewol}) and the often used range $50\div60$ has no model-independent justification. Instead of assuming a value (or range of values) of $N_k$ one may and should calculate it, given some assumptions about the reheating period. Let us illustrate this with the example of $\alpha$-attractor E-model (expressions for T- and P-models may be found in Appendix~\ref{app:TP-models}).

\begin{figure}[tbp]
\includegraphics[height=4.8cm]{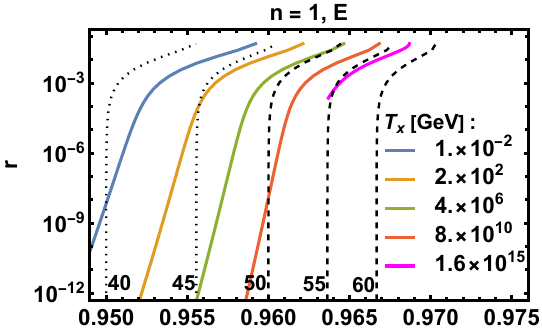}
\includegraphics[height=4.8cm,trim=17 0 0 0,clip]{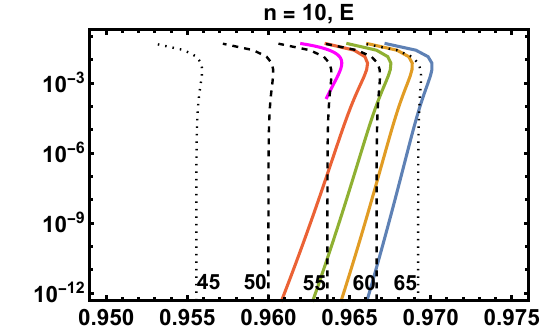}\\[6pt]
\includegraphics[height=4.8cm]{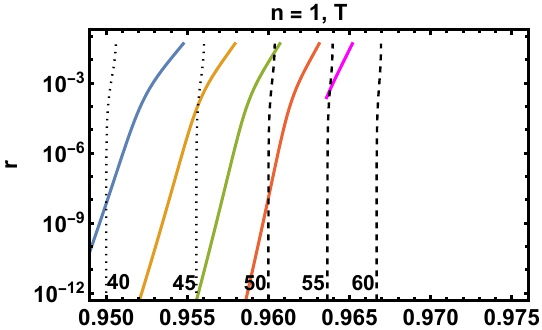}
\includegraphics[height=4.8cm,trim=17 0 0 0,clip]{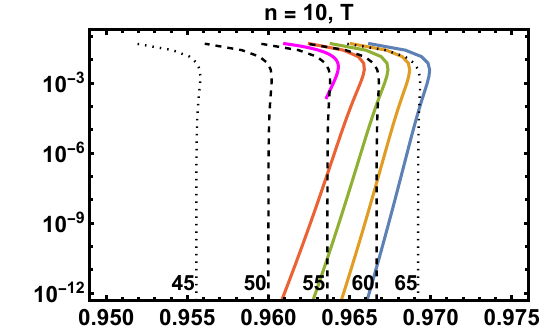}\\[6pt]
\includegraphics[height=5.35cm]{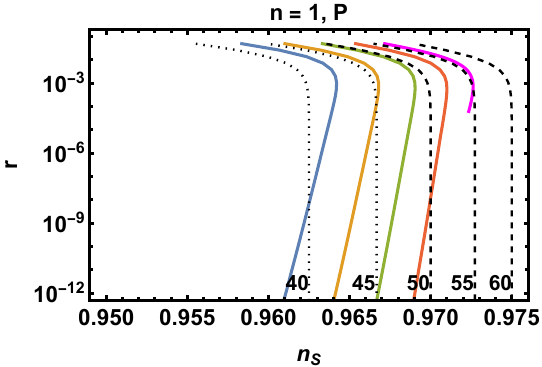}
\includegraphics[height=5.35cm,trim=17 0 0 0,clip]{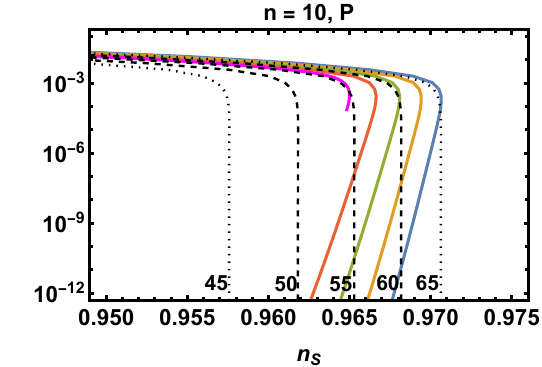}
  \caption{Correlation between $n_s$ and $r$ for three $\alpha$-attractor models (E, T, and P) and two values of the exponent $n$ (1 and 10). Colorful curves correspond to fixed values of the reheat temperature $\Tx$, while black curves represent fixed values of the number of e-folds $N_k$.}
  \label{fig:ns-r_Tx-Nk}
\end{figure}

The value of $\phi_{\rm end}$ at the end of (slow-roll) inflation can be estimated by the conditions for the slow-roll parameters $\epsilon =1$ or $|\eta |=1$, whichever is reached earlier. In the case of E-models, for the end of inflation defined by $\epsilon =1$, one gets: 
\begin{equation}
\label{eq:phi_end}
\phi_{\rm end }
=
\Mp \,\frac{4\sqrt{2}\,n}{(\xi-1)\sqrt{r}}\,
\ln\left(1+\frac{(\xi-1)\sqrt{r}}{4}\right)
\,.
\end{equation}
The number of e-folds, $N_k$, is expressible via $\phi_k$ and $\phi_{\rm end}$: 
\begin{equation}
   N_k =-\frac{1}{\Mp^2}\int_{\phi_k}^{\,\phi_{\rm end}}\frac{V(\phi)}{\partial_\phi V(\phi)}\,{\rm d}\phi\,.
   \label{eq:Nk-integral}
\end{equation}
Thus, using eqs.~\eqref{eq:phi_k} and \eqref{eq:phi_end}, one finds that $N_k$ is determined by the observables $n_s$ and $r$ as follows:
\begin{equation}
N_k
=
\frac{4n(4-\sqrt{r})}{r(\xi-1)}
-\frac{16n}{r(\xi-1)^2}
\,\ln\left(\frac{4\xi}{4+\sqrt{r}(\xi-1)}\right)
\,,
\label{eq:Nk}
\end{equation}
with $\xi$ given by eq.~\eqref{eq:xi}. Some curves of constant $N_{k_*}$ in the $(n_s(k_*),r(k_*))$
plane are shown for several models in Figure~\ref{fig:ns-r_Tx-Nk} (black dashed and dotted curves).

For each considered model the number of e-folds $N_{k_*}$ may be calculated for each pair of values of the CMB observables $n_s(k_*)$ and $r(k_*)$. However, not every pair of such values is consistent with the model of inflation
used for calculating $N_{k_*}$ and the assumed description of the reheating
process. One of the reasons is related to the model independent bounds on the reheating temperature which characterizes the transition from the reheating era to the radiation domination era. In our approach this temperature is a function of the CBM observables and the bounds on it constrain their acceptable range. 

The observables are measured by the Planck and other experiments for the comoving wavenumber
$k=k_* = a_{k_*}H_{k_*}$, with $a_{k_*}$ and $H_{k_*}$ denoting the scale factor and the Hubble scale at the moment of exit beyond the horizon of the mode with  comoving wavenumber $k_*$. One has an identity
\begin{equation}
   0 = \ln\left(\frac{k_*}{a_{k_*} H_{k_*}}\right)=\ln\left(\frac{a_{\rm end}}{a_{k_*}}\frac{a_{\rm re}}{a_{\rm end}}\frac{a_0}{a_{\rm re}}\frac{k_*}{a_0 H_{k_*}}\right),
    \label{eq:k=aH}
\end{equation}
where $a_{\rm end}$ and $a_{\rm re}$ are scale factors at the end of inflation and at the completion of reheating. The first three factors under the logarithm on the r.h.s.~of the above formula correspond to three regions in Figure~\ref{fig:aHinvewol}: inflation, reheating and standard evolution after reheating, respectively. It is a consistency relation reflecting the fact that the observed perturbation left the (event) horizon at the scale factor $a_{k_*}$ and reentered the (particle) horizon when the scale factor had some known value (depending of the choice of the pivot scale $k_*$).
The number of e-folds $N_{k_*}=\ln\left(\frac{a_{\rm end}}{a_{k*}}\right)$ from the horizon exit of the mode $k_*$ to the end of inflation has already been given in eq.~\eqref{eq:Nk-integral} in terms of the inflaton  potential parameters and the value  of the inflaton $\phi_{\rm end}$ at the end of inflation, expressed by   
the CMB observables $n_s(k_*)$, $r(k_*)$ and $A_s(k_*)$. 
Regarding the reheating period, we adopt the standard description by the Boltzmann equation of the energy density transfer to radiation, with the inflaton dissipation rate  $\Gamma$ as a free parameter. 
Some non-perturbative effects are also taken into account.
We trade $\Gamma$ for the reheating temperature $T_{\rm re}$. It is then clear that for a given model of inflation the identity eq.~\eqref{eq:k=aH}  gives us the reheating temperature (or $\Gamma$) fixed in terms of  the observables
$A_s(k_*)$, $n_s(k_*)$ and $r(k_*)$. Conversely, some well motivated bounds on the reheating temperature, as we see later, can be translated into very severe tests of inflaton models by the CMB data.

We describe now the details of the above outlined procedure. 
We start with the calculations, ignoring non-perturbative effects, the influence of which we will analyze later.
Three approximations will be used in deriving analytical expressions describing the reheating process:
\begin{itemize}
\item
Inflaton potential during the reheating will be approximated by the appropriate monomial of the infalton field i.e.~the first term of the Taylor expansion of the full potential \eqref{eq:alpha-attractors}.
\item
Contribution of the radiation energy density to the Hubble parameter during the reheating will be neglected (such approximation is very accurate except the very late stage of reheating).
\item
Contribution of the inflaton energy density after the reheating process will be neglected.
\end{itemize} 
The accuracy of the results obtained using these approximations will be discussed in Appendix~\ref{app:approx}.

We define the reheating temperature $T_{\rm re}\equiv \Tx$ as the temperature 
of radiation at the moment when $\rho_\phi = \rho_R$, where $\rho_\phi$ and $\rho_R$ are the inflaton and radiation energy densities, respectively, that is, the time or the temperature at which the energy densities cross each other.
The value of $\Tx$  enters into eq.~\eqref{eq:k=aH}  in the ratios $\left(\frac{a_{\rm \times}}{a_{\rm end}}\right)$  and $\left(\frac{a_0}{a_ \times}\right)$ (now $a_{\rm re}\equiv a_\times)$. 
For the latter, using the last of the listed above approximations, we take 
the standard form:
\begin{equation}
\label{eq:axa0}
    \begin{aligned}
        \frac{a_{\times}}{a_0}&=\left(\frac{43}{11g_{s*}}\right)^{1/3}\frac{T_0}{\Tx}\,,
    \end{aligned}    
    \end{equation}
where $T_0$ is the present temperature of the universe.    
The ratio $\left(\frac{a_\times}{ a_{\rm end}}\right)$ requires more attention.
\bigskip
Classical reheating is described by the following set of Boltzmann equations\footnote{We assume $\Gamma$ to remain constant during the entire reheating process.}:
\begin{eqnarray}
\dot{\rho_\phi} &=& - 3(1+w)H\rho_\phi - \Gamma \rho_\phi
\label{rho_phi_dot}
\,,\\
\dot{\rho_R} &=& - 4H\rho_R + \Gamma \rho_\phi
\label{rho_R_dot}
\,,\\
H^2 &=& \frac{\rho_\phi + \rho_R}{3\Mp^2}
\,,
\label{H_with_rho_R}
\end{eqnarray}
where dots denote derivatives with respect to the cosmic time.
$\rho_\phi$ is the total (potential + kinetic) energy of the inflaton averaged over one period of its oscillations and $w$ is the accordingly averaged inflaton equation of state parameter (during one period the temporary value of the inflaton equation of state parameter changes in the range $[-1, +1]$). For a monomial potential $V={\rm const}\cdot\phi^{2n}$ one gets $w=(n-1)/(n+1)$. In our analytical calculations we assume that the inflaton potential {\it during oscillations} may be approximated by a monomial one so we use 
\begin{equation} \label{w}
w\approx \frac{n-1}{n+1}\;.
\end{equation}
Such approximation is very good for $\phi\ll M_{\rm P}$ because Taylor expansion of each potential \eqref{eq:alpha-attractors} starts with the monomial of order $2n$ and higher terms are suppressed by additional powers of $(\phi/M_{\rm P})^2$. Some corrections appear at very early stage of oscillations but they soon become negligible because the amplitude of oscillations decreases quite fast in the expanding universe, especially for bigger values of $n$ (see Appendix~\ref{app:approx}).

It occurs that it is more convenient to analyze the set of equations \eqref{rho_phi_dot}--\eqref{H_with_rho_R} using the cosmic scale factor, $a$, instead of the cosmic time as the independent variable. 
We will apply also the usual approximation consisting in neglecting the radiation contribution to the total energy density during the reheating process, i.e.~$\rho_R\ll\rho_\phi$, resulting in the 
following approximate expression for the Hubble parameter 
(the impact of this approximation on our numerical results is discussed in Appendix~\ref{app:approx}):
\begin{equation}
H^2 \approx \frac{\rho_\phi}{3\Mp^2}\,.
\label{H_approx}
\end{equation}
This way equations \eqref{rho_phi_dot} and \eqref{rho_R_dot} with $H$ given by \eqref{H_approx} may be rewritten in the following form:
\begin{eqnarray}
a\rho_\phi^\prime &=& -3(1+w)\rho_\phi - \sqrt{3}\Mp\Gamma\sqrt{\rho_\phi}
\label{rho_phi_prime}
\,,\\
a\rho_R^\prime &=& -4\rho_R + \sqrt{3}\Mp\Gamma\sqrt{\rho_\phi}
\label{rho_R_prime}
\,,
\end{eqnarray}
where primes denote derivatives with respect to the cosmic scale factor $a$.
The above set of equations may be solved analytically. The solution reads
\footnote{
These results reduce to those from standard textbooks (see e.g.~\cite{Kolb:1990vq}) if $\Gamma$ is set to zero in \eqref{gamma_def}, but explicit $\Gamma$ is kept in \eqref{rho_R_sol}. This corresponds to neglecting $\Gamma$ term in \eqref{rho_phi_prime} but keeping it in \eqref{rho_R_prime}.}
\begin{eqnarray}
\rho_\phi
&=&
\rho_{\rm end}\left[(1+\gamma)\left(\frac{a}{a_{\rm end}}\right)^{-\frac32(1+w)}-\gamma\right]^2
\label{rho_phi_sol}
,\\
\rho_R
&=&
\rho_{\rm end}\,\frac{\Gamma}{H_{\rm end}}\left[\frac{2(1+\gamma)}{5-3w}\left(\left(\frac{a}{a_{\rm end}}\right)^{-\frac32(1+w)}-\left(\frac{a}{a_{\rm end}}\right)^{-4}\right)-\frac{\gamma}{4}\left(1-\left(\frac{a}{a_{\rm end}}\right)^{-4}\right)\right]
,\qquad
\label{rho_R_sol}
\end{eqnarray}
where
\begin{equation}
\gamma=\frac{\Gamma}{3(1+w)H_{\rm end}}
\,,
\label{gamma_def}
\end{equation}
and we used obvious initial conditions at the beginning of reheating (identified with the end of inflation): $\rho_R(a_{\rm end})=0$, $\rho_\phi(a_{\rm end})=\rho_{\rm end}$. For a given model energy density at the end of inflation may be calculated in terms of the CBM observables. In the case of E-models it reads
\begin{equation}
\rho_{\rm end } =
2\pi^2\Mp^4\,r\,A_s\,
\left(\frac{\xi\sqrt{r}}{(\xi-1)\sqrt{r}+4}\right)^{2n}
\,.
\end{equation}

We define the end of reheating as the moment when the cosmic scale factor is equal $a_\times$ for which $\rho_\phi(a_\times)=\rho_R(a_\times)$.
It is possible to calculate $a_\times$ after neglecting terms proportional to $(a/a_{\rm end})^{-4}$ in \eqref{rho_R_sol}.
We obtain
\begin{equation}
\frac{a_\times}{a_{\rm end}}
\approx
\left[
\frac{\gamma\left(16+3(1+w)\sqrt{9-3w}\right)}{2(1+\gamma)(5-3w)}
\right]^{-\frac{2}{3(1+w)}}\,.
\label{a_times}
\end{equation}

Energy density at $a_\times$ equals $\rho_\times=\rho_\phi(a_\times)+\rho_R(a_\times)=2\rho_\phi(a_\times)$. Substituting $a_\times$ given by \eqref{a_times} into \eqref{rho_phi_sol} we obtain the following simple result
\begin{equation}
\rho_\times=\frac{\rho_{\rm end}}{2}\left[3(1+w)\gamma\,\frac{2+\sqrt{9-3w}}{5-3w}\right]^2
\,.
\label{rho_times}
\end{equation}
From the equality $\rho_\phi(a_\times)=\rho_R(a_\times)$ one may calculate the reheating temperature as a function of $\gamma$ (defined in eq.~\eqref{gamma_def})
\begin{equation}
\Tx^4
=
\frac{30}{g_*\pi^2}\,\frac{\rho_\times}{2}
=
\frac{30}{g_*\pi^2}\,\rho_{\rm end}
\left[\frac32(1+w)\gamma\,\frac{2+\sqrt{9-3w}}{5-3w}\right]^2
\,,
\label{T_times}
\end{equation}
which may be easily inverted:
\begin{equation}
{\gamma}
=
\frac{\Tx^2}{\sqrt{\rho_{\rm end}}}\,
\sqrt{\frac{2g_*}{15}}\frac{\pi(5-3w)}{3(1+w)(2+\sqrt{9-3w})}
\,.
\label{gamma}
\end{equation}
In calculations with fixed reheating temperature $\Tx$ one should replace $\gamma$ with the above function of $\Tx$.

The relation between the reheating temperature $T_\times$ defined above eq.~\eqref{eq:axa0} and the frequently used ``decay-completion'' temperature $T_{\Gamma=H}$ at which $\Gamma\sim H$ is given in formula \eqref{eq:TxTGamma=H}.
Notice also that using our procedure there is no need to define and use the  effective equation of state parameter $w_{\rm eff}$, often utilized in the literature 
to parameterize the final effect caused by evolution of the system including inflaton and radiation
\footnote{Results \eqref{a_times} and \eqref{rho_times} may be used to calculate $w_{\rm eff}$ which is defined by
$
{\rho(a_1)}/{\rho(a_2)}
=
\left({a_1}/{a_2}\right)^{-3(1+w_{\rm eff})}
$.
In the case of reheating we have $a_1=a_\times$ and $a_2=a_{\rm end}$. The expression for effective $w$ reads
\begin{equation}
w_{\rm eff}
=
\frac{\displaystyle
\ln\left(\frac{3(1+w)\left(2+\sqrt{9-3w}\right)}{\sqrt{2}(5-3w)}\,\gamma\right)}
{\displaystyle
\ln\left(\frac{\left(16+3(1+w)\sqrt{9-3w}\right)}{2(5-3w)}\,\frac{\gamma}{1+\gamma}\right)}-1
\,.\nonumber
\end{equation}
}.

We checked, by solving numerically the Boltzmann equations \eqref{rho_phi_dot}--\eqref{H_with_rho_R} for several sample models and parameter sets,
that accuracy of our analytical results \eqref{a_times}--\eqref{T_times} is very good when two conditions are met. First: $w<\frac53$ which is always the case in the $\alpha$-attractor models considered in this paper. Second: $\Gamma$ is not much bigger than about $0.1 H_{\rm end}$. The results for bigger values of the ratio $\Gamma/H_{\rm end}$ becomes gradually less precise (see Appendix~\ref{app:approx} for more quantitative discussion). However, for $\Gamma$ of the order of or bigger than $H_{\rm end}$ we are close to the instant reheating limit for which the whole analysis of the reheating process is no longer needed. The instant reheating limit comes down to the assumption that reheating is so rapid that no details of it matter. Moreover, very rapid reheating corresponds to smallest possible values of $r$ for given $T_\times$. Typically such values of $r$ are many orders of magnitude below bounds from present or planed experiments. Later we will discuss this in more detail.

The result \eqref{a_times} is important for us because the ratio $a_\times/a_{\rm end}$, after substituting $\gamma$ given by \eqref{gamma}, may be directly used in the relation~\eqref{eq:k=aH} (with $a_{\rm re}\equiv a_\times$). 
Finally, using the Friedman equation and definitions of the parameters $r$ and $A_s$, one obtains the following expression for $H_k$
\begin{equation}
\label{eq:H_k_as_fun_of_inflaton_params}
    H_k = \frac{\pi}{\sqrt{2}} \Mp \sqrt{rA_s}\,.
\end{equation}
This completes our task of relating the values of the observables $n_s, r, A_s$ to a value of $\Tx$ in an $\alpha$-attractor inflation potential 
with a given exponent $n$. This is an important prediction  due to the following: Firstly, there are model independent bounds on the reheat temperature of the universe
\begin{equation} 
10\,\MeV\lesssim \Tx \lesssim 2\cdot10^{15}\,\GeV\,.
\label{eq:Tx-bounds}
\end{equation}
The lower bound is the BBN energy scale while the upper one 
follows from the fact that the consistency condition \eqref{eq:k=aH} with too high reheat temperatures has no physically meaningful solutions.
Increasing $T_\times$ results in bigger number of e-folds during the RD period so fulfilling condition \eqref{eq:k=aH} requires smaller number of e-folds during the reheating period. Exceeding the upper limit on $T_\times$ would require a negative number of e-folds during reheating, which of course can not be realized.

The above bounds are not precise. The highest possible $T_\times$ for which consistency condition \eqref{eq:k=aH} have solutions depends slightly on a considered model. We have checked numerically that for the models analyzed in this work differences in maximal $T_\times$ are not bigger than by a factor of about 2. Similarly, some analyses show that the BBN energy scale may be slightly lower than 10 MeV. However, changing the maximal and/or minimal value of $T_\times$ by a factor of 2 would have very little effect on the conclusions drawn from our calculations. Typical change of $n_s$ caused by changing $T_\times$ by a factor of 2 (with $r$ fixed) is of order $10^{-6}$.

The bounds \eqref{eq:Tx-bounds} on $T_\times$ provide strong tests via the CMB data on the models of inflation. Moreover, that explicit link of the CMB data to the reheat temperature may have important particle physics implications. Several theoretical consistency conditions and new physics ball-park values which are often conveyed through reheat temperatures  will readily find observable quantities (in a given model of inflation) like spectral index, tensor-to-scalar ratio, etc.~which either will be constrained or will be tested further with upcoming precision CMB measurements. 
Since at present only the  upper bound on $r$ is known experimentally,  it is interesting to illustrate the above  results in the plane ($n_s$, $r$) as a function $r(n_s)= f(n, A_s, n_s, \Tx)$, for 
several fixed values of the exponent $n$. First, we compare our approach to the selection of the acceptable inflationary models based on the reference to the reheating temperature $\Tx$ with the one using an ansatz for the e-fold number $N_k$ to be in the range $50\div60$. 
In Figure~\ref{fig:ns-r_Tx-Nk} two sets of curves in the ($n_s$, $r$) plane are compared: colored curves with fixed $\Tx$, $r(n_s)= f(n, A_s, n_s, \Tx)$, and black ones with fixed $N_k$, $r(n_s)= f(n, A_s, n_s, N_k)$. 
For $r\gtrsim10^{-3}$ the two sets of curves have similar $r$-dependence, with the fixed temperature curves consistent with some fixed $N_k$, but not always in the range $50\div60$.
The characters of those two sets of curves as functions of $r$  
are quite different for smaller values of $r$. A given model may be self-consistent only if it predicts a point on the ($n_s$, $r$) plane between the outermost colored curves (with very small corrections discussed in the previous paragraph) 
because only then the consistency condition \eqref{eq:k=aH} and the model independent bound on the reheat temperature \eqref{eq:Tx-bounds} are fulfilled.

The curves in Figure~\ref{fig:ns-r_Tx-Nk} for the highest considered $T_\times$ are quite short. Their lower end-points at $r$ slightly smaller than $10^{-4}$ correspond to instantaneous reheating. For smaller values of $r$ there are no solutions of the consistency condition \eqref{eq:k=aH}. Analogous lower limits on $r$ do exist for all temperatures $T_\times$. However, they scale approximately as the fourth power of $T_\times$ so for example the lower allowed $r$ for $T_\times=8\cdot10^8$\,GeV ($T_\times=10$\,MeV) is of order $10^{-21}$ ($10^{-73}$) i.e.~far outside the range of $r$ shown in all our figures. Such scaling follows from the fact that $r$ is approximately proportional to the forth power of the energy scale of the inflaton potential (see eq.~\eqref{eq:Lambda_inf}) which in the case of instantaneous reheating is simply related to the forth power of the reheating temperature (inflaton energy at the end of inflation is instantaneously transformed into radiation energy which is proportional to $T_\times^4$).

As mentioned below eq.~\eqref{gamma}, our analytical results become less accurate when the reheating is very rapid. However, the above discussion shows that this corresponds to $r$ being close to its minimal possible value (for given $T_\times
$). Such small values of $r$ are outside ranges shown in our figures (with the only exception of $T_\times=1.6\cdot10^{15}$\,GeV in Figure~\ref{fig:ns-r_Tx-Nk}).

The strong dependence of the lower possible value of $r$ on $T_\times$ is related to the upper bound \eqref{eq:Tx-bounds} on $T_\times$. Allowed range of $r$ shrinks very fast with $T_\times$ increasing above the highest $T_\times$ shown in our plots, i.e.~above $1.6\cdot10^{15}$\,GeV. The highest possible $T_\times$ corresponds to such range shrunk to a point.

In many models the allowed region in the ($n_s$, $r$) plane has only some partial overlap with the region corresponding to the often used condition $50\le N_k\le 60$. 
Clearly, the approach based on well motivated bounds on $\Tx$ may give significantly different results compared to an (essentially arbitrary) ansatz for $N_k$.
Thus, in the following the dependence of the results on the reheating temperature (and not on $N_k$) will be investigated.

There are several interesting observations following from Figure~\ref{fig:ns-r_Tx-Nk}. 
Very narrow ranges of $n_s$ are predicted, that depend on the model and on $r$. 
In most of the considered cases there is quite strong qualitative change in the behaviour of the curves of constant $\Tx$ for some, depending on the model, value of $r$ in the range $10^{-4}\div10^{-2}$. 
For smaller $r$ the correlations between $n_s$ and $r$ are quite similar for E- and T-models for all considered values of $n$. Such correlations are different in the case of P-models, especially for $n=1$ or for relatively big values of $r$. 
For all models the spectral index $n_s$ decreases with decreasing $r$ for $r\lesssim10^{-4}$. This, together with experimental lower bounds on $n_s$, results in lower bounds on $r$. Such bounds are different for different models and strongly depend on the reheating temperature $T_\times$. For example, using the Planck results at 2$\sigma$ level in the case of E-model with $n=1$ we obtain $r\gtrsim0.004$ for $T_\times=200$\,GeV and $r\gtrsim10^{-12}$ for $T_\times=8\cdot10^{10}$\,GeV. The corresponding bounds on $r$ in the case of P-model are weaker by many orders of magnitude. Nevertheless, with the exception of very high $T_\times$, such smallest values of $r$ consistent with experimental bounds on $n_s$ are (much) bigger than values of $r$ corresponding to instantaneous reheating.

\begin{figure}[tbp]
    \includegraphics[height=4.6cm]{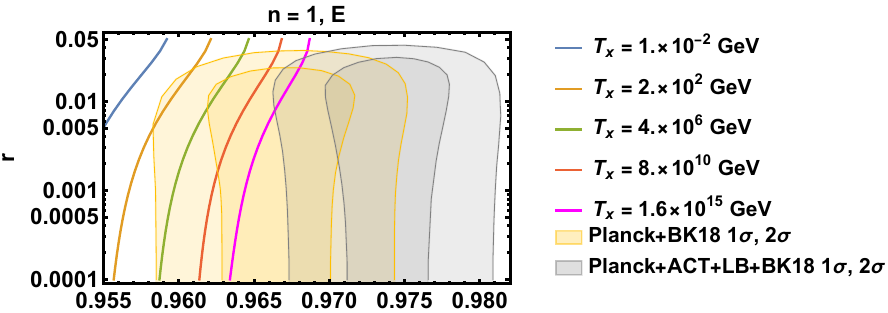} \\[6pt]
\includegraphics[height=4.55cm]{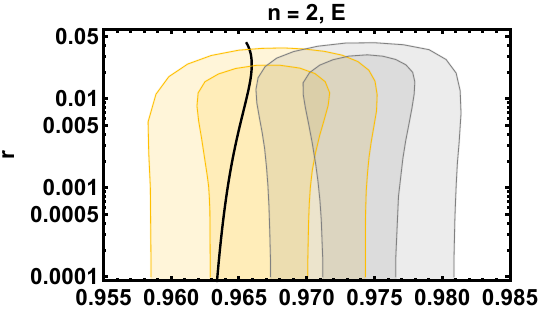} 
\includegraphics[height=4.55cm, trim=14 0 0 0,clip]{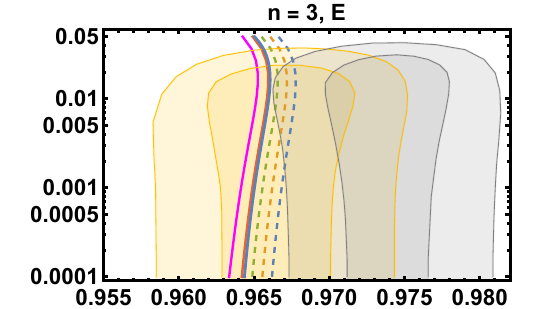} 
\\[6pt]
\includegraphics[height=5.15cm]{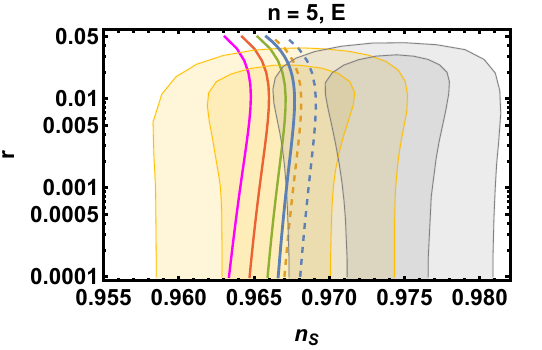} 
\includegraphics[height=5.15cm, trim=14 0 0 0,clip]{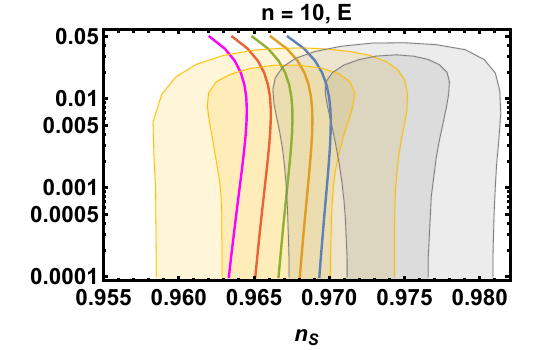}
\caption{Correlation between $n_s$ and $r$ as a function of the reheat temperature $\Tx$ in the case of E-model for several values of the exponent $n$. Results for $n=2$ do not depend on $\Tx$ and are represented by the black curve. The experimental $1\sigma$ and $2\sigma$ allowed regions from \textit{Planck}, \textit{BK18} and \textit{BAO} \cite{Planck:2018vyg,BICEP2:2018kqh} are shown. 
The dashed lines correspond to the results obtained without taking into account the non-perturbative fragmentation of the inflaton condensate.}
\label{fig:nsrPlotEmodel} 
\end{figure}

\begin{figure}[tbp]  
\includegraphics[height=4.55cm]{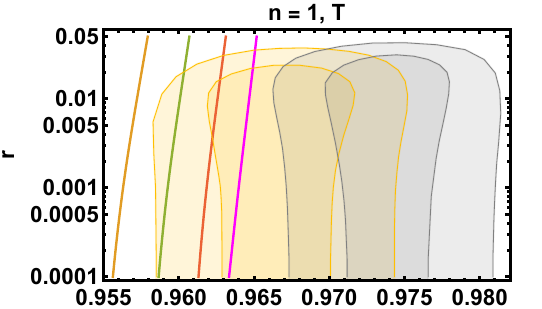} 
\includegraphics[height=4.55cm, trim=14 0 0 0,clip]{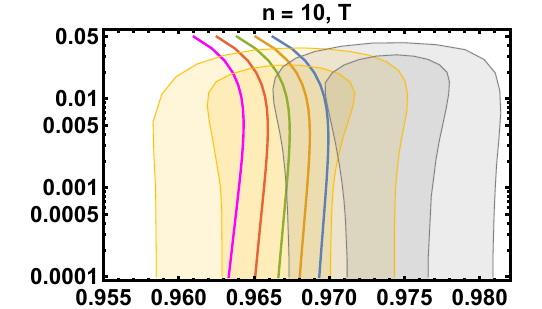} \\[6pt]
\includegraphics[height=5.15cm]{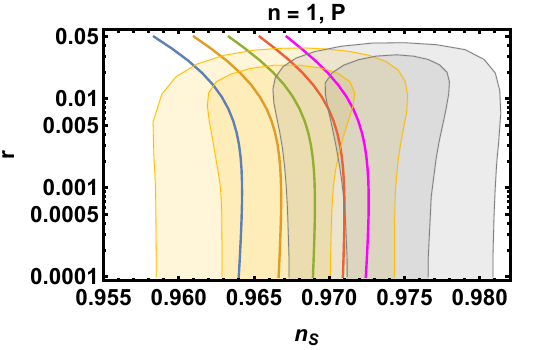}
\includegraphics[height=5.15cm, trim=14 0 0 0,clip]{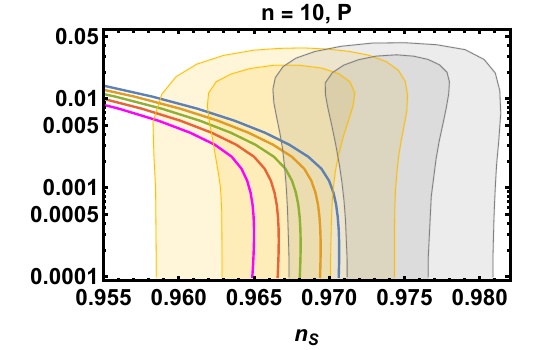} 
\caption{
Plots analogous to those in Figure~\ref{fig:nsrPlotEmodel} for T- and P-models with $n=1$ and $n=10$.}
\label{fig:nsrPlotTPmodels} 
\end{figure}

Results for $r\ge10^{-4}$ are presented in Figures~\ref{fig:nsrPlotEmodel} and \ref{fig:nsrPlotTPmodels} where curves of constant $\Tx$ are shown together with present experimental bounds.
The present $2\sigma$ contours of the Planck+BK18 
data differ quite strongly from the global fit including also ACT and DESI data 
(for a recent, detailed discussion  of experimental data and the differences between those experiments see e.g.~\cite{Ferreira:2025lrd,Ellis:2025zrf}). 
The latter fit has more discriminative power for the considered models
in the sense that the predictions of some models (for some values of $n$ and some ranges of $T_\times$) lie outside the experimental $2\sigma$ region. 
For example fit including ACT and DESI data disfavor: in the case of $n=1$ T-models for arbitrary $T_\times$, E-model with exception of very big $T_\times$ and $r$, P-model with low $T_\times$; for $n>2$ all models with relatively high $T_\times$. On the other hand, only E- and T-models with $n=1$ and very low $T_\times$ are excluded by the Planck+BK18 data at $2\sigma$ level. 
The global fit shows some preference for P-model with $n=1$ and for high reheat temperature $T_\times$. It can also be reconciled with E and T-models with large values of $n$ and low reheat temperature. 
Actually, we see that in the considered models there is the upper bound on $n_s$ around 0.97, to be compared with the $2\sigma$ lower bound of the global fit at $\sim0.965$. 
One can also ask how constraining are the present CMB data for the reheating temperature. Given the large difference between the Planck data and the global fit one concludes that the whole range of temperatures is acceptable. Taken Planck and the global fit separately, some constraints on the reheating temperatures can be seen in the figures.

Furthermore, it is worth noting the change of ordering
of the $\Tx = {\rm const}$ curves from large to small values, as a function of $n_s$. This is related to the fact that $w<\frac13$ for $n=1$ while $w>\frac13$ for $n>2$. 
If $w<\frac13$ during reheating, lower $T_\times$ (for fixed value of $r$) requires smaller number of e-folds $N_k$ during inflation after the pivot scales leaves the Hubble horizon.
This may be qualitatively seen by comparing blue and red lines in Figure \ref{fig:aHinvewol}. 
The effect is opposite for $w>\frac13$ (green line in Figure \ref{fig:aHinvewol}). 
Different dependence of $N_k$ on $T_\times$ results in different dependence of $n_s$.
The case of $n=2$ is a special one because it is described by $w=\frac13$ so the the number of e-folds $N_k$ does not depend on $T_\times$ (for fixed $r$). As a result the function $r(n_s)$ does not depend on $\Tx$ 
(the black curve in Figure~\ref{fig:nsrPlotEmodel}).

\begin{figure}[tbp]
   \includegraphics[height=4.49cm]{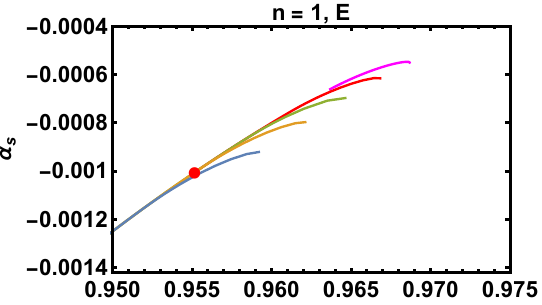} 
\includegraphics[height=4.49cm, trim=14 0 0 0,clip]{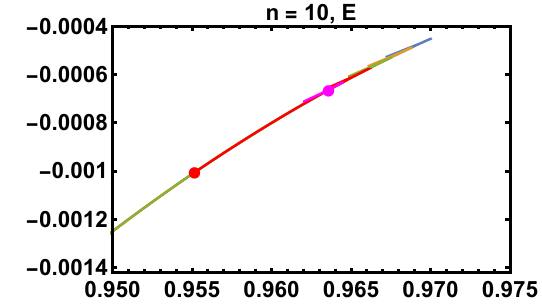} \\[6pt]
\includegraphics[height=5.0cm]{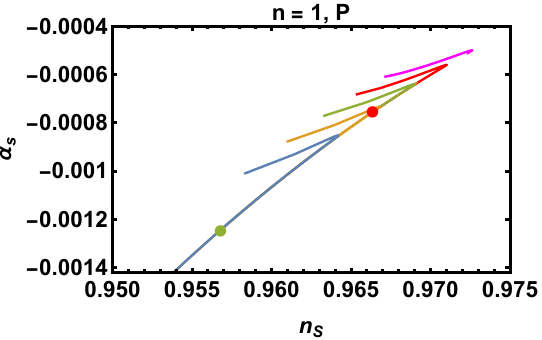} 
\includegraphics[height=5.0cm, trim=14 0 0 0,clip]{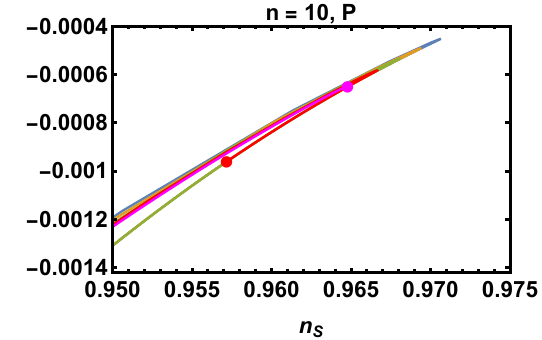} 
\caption{\label{fig:alfasrPlotEPmodels} Correlation between scalar spectral index $n_s$ and its running $\alpha_s$, following from eq.~\eqref{eq:alphaS_n_r_ns}, in the case of E- and P-model for two values of the exponent $n=1,10$ for fixed reheat tempearature $\Tx$ (color coding as in Figures~\ref{fig:ns-r_Tx-Nk}-\ref{fig:nsrPlotTPmodels}). As the curves in different colours partially overlap, the end of some of them is denoted by the dot of the same clour. The absence of a dot in a given color means that the corresponding curve ends outside the plot.}
\end{figure}

Results for $n=3$ and $n=5$ shown on two panels in Figure \ref{fig:nsrPlotEmodel}
require more explanation 
because they are examples of models for which non-perturbative effects are important. 
Dashed color curves show the correlation between $r$ and $n_s$ with such effects neglected. Taking into account non-perturbative fragmentation of the inflaton condensate changes results for some range of the reheating temperatures \cite{Garcia:2023eol,Garcia:2023dyf,Ellis:2025zrf}. This happens if fragmentation takes place before the end of reheating 
because after fragmentation the inflaton equation of state parameter $w$ changes its value from $(n-1)/(n+1)$ to $w=\frac13$ 
(number of e-folds after which this happens for several monomial potentials was estimated in \cite{Garcia:2023dyf}). 
Then, in analogy to the $n=2$ case, the evolution of the universe no longer depends on the value of $T_\times$. As a result the dependence $r(n_s)$ is the same for all $T_\times$ smaller than some critical temperature. This critical temperature corresponds to the situation when fragmentation happens just at the end of reheating. The resulting $r(n_s)$ is shown by the rightmost color curves on panels for $n=3$ and $n=5$. With non-perturbative fragmentation of the inflaton condensate taken into account all dashed curves must be moved to the left and coincide with the described above critical curve.

Inflaton fragmentation does not influence any other panel in Figures \ref{fig:ns-r_Tx-Nk}--\ref{fig:alfasrPlotEPmodels}. The reasons are the following:\\
-- there is no fragmentation in quadratic potential, i.e.~for $n=1$;\\
-- fragmentation in quartic potential, i.e.~for $n=2$, does not change $w$ (which also before fragmentation equals $\frac13$);\\
-- fragmentation does not change results for $n>7$ because it would happen after the end of reheating even for smallest considered $T_\times$ \cite{Garcia:2023dyf}.

There are several experiments, \textit{LiteBIRD}, \textit{CMB-S4}, and \textit{SO} \cite{LiteBIRD:2022cnt,CMB-S4:2016ple,SimonsObservatory:2018koc}, which are expected to provide stronger bounds on the CMB observables. Sensitivity for $r$ is predicted to be more or less an order of magnitude better than that of present data. The accuracy of determining the spectral index $n_s$ should also be better. The 2$\sigma$ range of $n_s$ in the case of \textit{CMB-S4} and \textit{SO} is predicted to be of about 0.01. This could lead to tighter constraints on models, but the exact results will depend on the obtained central values for $n_s$ and $r$ (or the upper bound in the case of $r$).

In Figure~\ref{fig:alfasrPlotEPmodels} we show the curves of constant $\Tx$ in the plane ($n_s$, $\alpha_s$) for E- and P-models. The bounds on the reheating temperature give the lower bound on the absolute value of $\alpha_s$ around $5\times10^{-4}$,  independently of the considered model. 
Comparing Figure~\ref{fig:alfasrPlotEPmodels} with Figure~\ref{fig:ns-r_Tx-Nk} we see that the maximum absolute values of $\alpha_s$ allowed by the $\Tx$ bounds are increasing with decreasing values of $r$. This may be understood from the expressions for $\alpha_s$ in terms of the CMB observables. Substituting $\xi$ given by eq.~\eqref{eq:xi} into eqs.~\eqref{eq:alphaS-E} and \eqref{eq:alphaS-P} we get the following simple expressions
 \begin{equation}
     \alpha_s =
     \begin{cases}
   -\displaystyle\frac{\left(8(1-n_s)-r\right)\left(n\left(8(1-n_s)-r\right)-r\right)}{128n}
 &\quad\text{\qquad for E-model,}\\[8pt]
   -\displaystyle\frac{64\left(1-n_s\right)^2+n\left(8\left(1-n_s\right)-r\right)^2+r^2}{64(2n+1)}
&\quad\text{\qquad for P-model.}
\end{cases}
\label{eq:alphaS_n_r_ns}
 \end{equation}
The dependence on $r$ is very weak for $r\ll(1-n_s)$ with the following simple limits:
\begin{equation}
    \lim_{r\to0}\alpha_s
    =
    \begin{cases}
    -\displaystyle\frac12\left(1-n_s\right)^2
 &\qquad\text{\qquad for E-model,}
 \\[6pt]
  -\displaystyle\frac{n+1}{2n+1}\left(1-n_s\right)^2
&\qquad\text{\qquad for P-model.}
\end{cases}
\label{eq:alphaS_limit}
\end{equation}
In the case of E-models $\alpha_s$ for small $r$ depends only on $n_s$ and does not depend on the exponent $n$. This is visible on upper panels of Figure~\ref{fig:alfasrPlotEPmodels} where the behavior of $\alpha_s$ for small $r$ (corresponding to small $n_s$) is almost the same for $n=1$ and $n=10$. Situation in P-models is somewhat different. The running of the spectral index does depend on $n$. The first factor in the lower line of the r.h.s.~of \eqref{eq:alphaS_limit} approaches $-\frac12$  present in upper line of \eqref{eq:alphaS_limit} in the limit of large $n$. One can see in Figure~\ref{fig:alfasrPlotEPmodels} that the behavior of $\alpha_s$ for small $r$ in P-models with $n=10$ is similar to that in E-models. The biggest difference is between P-model with $n=1$ and E-models (with arbitrary $n$). Comparing two lines in eq.~\eqref{eq:alphaS_limit} one can see that $\left|\alpha_s\right|$ in such P-model is bigger by the factor of $\frac43$. The present $2\sigma$ lower bound on $n_s$ may be translated into upper bound on $|\alpha_s|$. Depending on the model this bound is between $8\cdot10^{-5}$ and $1.1\cdot10^{-4}$ with the highest value for P-model with $n=1$.
These numbers give us some intuition about the required sensitivity of the $\alpha_s$ measurements for being useful as a supplement to $n_s$ in discriminating between different models of inflation, if $r$ turns out to be very small.
Determination of $\alpha_s$ with accuracy substantially better than $10^{-3}$ would be desirable. 
Present experimental bounds are much weaker so using $\alpha_s$ as the fourth input observable would not result in further constraints on the parameters of inflationary models.
Experimental sensitivity slightly better than $10^{-3}$ may be reached in some planned experiments \cite{SPHEREx:2014bgr} but obtaining more precise experimental constraints in near future seems to be problematic (see e.g.~\cite{Bahr-Kalus:2022prj} and references therein).


\medskip
\section{\boldmath $\Tx$ sensitive dark matter production}
\label{sec:DM}

\begin{figure}[t]\centering  
        \centering
              \includegraphics[width=0.5\textwidth]{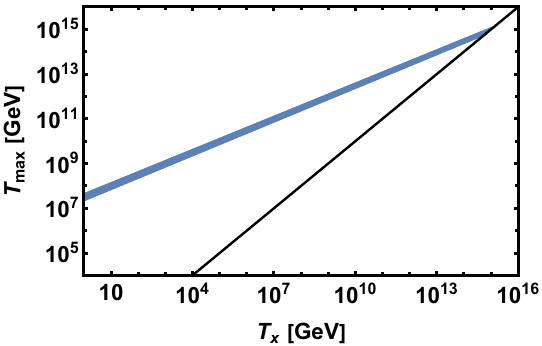}
    \caption{\label{fig:tmaxvstrhplot} The blue line shows $T_{\rm max}$ vs $\Tx$. Its thickness reflects some (very weak) dependence on the parameters $\Gamma$ and $w$.
The maximal spread of $T_{\rm max}$ for a given value of $\Tx$ for the range of parameters explored in this work is less than a factor of about 1.5. 
 The black line denotes $T_{\rm max}= \Tx$.}
\end{figure}

In this section we discuss the reheating temperature as a portal to link some DM production mechanisms to the CMB observables. One obtains then an interesting consistency check between theories of inflation and DM production. That happens when DM production is sensitive to the reheating
temperature $\Tx$, as is the case in some freeze-in models. We consider the DM production in a process of annihilation of some particles (e.g.~SM
particles) in thermal bath during the reheating process and after reaching the temperature $\Tx$. 
To consider both periods on equal footing, it is useful to recall that the reheating period is also characterized by some temperature usually denoted by $T_{\rm max}$, That temperature is calculable in terms of the $\Tx$  so that overall, given a DM production model, requiring $\Omega_{\rm DM} h^2=0.12$ indeed fixes the value of $\Tx$. 
The temperature
$T_{\rm max}$ may be easily obtained from maximization of \eqref{rho_R_sol} -- of course this time NOT neglecting any terms. Value of the cosmic scale factor for which $\rho_R$ is maximal equals:
\begin{equation}
a_{\rm max} = a_{\rm end}
\left(\frac{8+3(1+w)\gamma}{3(1+w)(1+\gamma)}\right)^\frac{2}{5-3w}
\end{equation}
This should be used in expression for $\rho_R$ given by \eqref{rho_R_sol} with $a$ replaced with the above $a_{\rm max}$.
The results are shown in Figure~\ref{fig:tmaxvstrhplot}, for a large range of values of $\Gamma$ and several values of the inflaton $w$. The dependence on those parameters is very weak, it corresponds to the thickness of the blue line.

If the DM particles are weakly enough coupled to the thermal bath they never reach thermal equilibrium with it but nevertheless they may reach the observed DM relic abundance in that freeze-in process. The necessary condition for the $\Tx$ dependence of that production process is that in the accessible range of energies (determined
by the value of $\Tx$) its amplitude can effectively be derived from a
higher dimension operator, ${\cal L}\sim {\cal O}_{4+d}/\Lambda^{d}_{\rm DM}$, (we take $d$ to be integer)\footnote{Parameter $d$ can also be fractional motivated from conformal field theory operators where the DM sector resides in CFT, investigated recently in refs.~\cite{Hong:2019nwd,Hong:2022gzo,Csaki:2021gfm} with intriguing consequences in DM experiments.} where $\Lambda_{\rm DM}$  is some physical cut-off scale to the effective theory.
It is clear on dimensional grounds  that, for production of relativistic particles considered here, the thermal average cross-section, 
relevant for the Boltzmann equation describing the production yield, reads
\begin{equation}
    \langle\sigma v\rangle\sim \frac{T^{2(d-1)}}{\Lambda^{2d}_{DM}}.
\label{eq:crosssection}    
\end{equation}

\begin{figure}[tbp]  
\includegraphics[height=5.0cm]{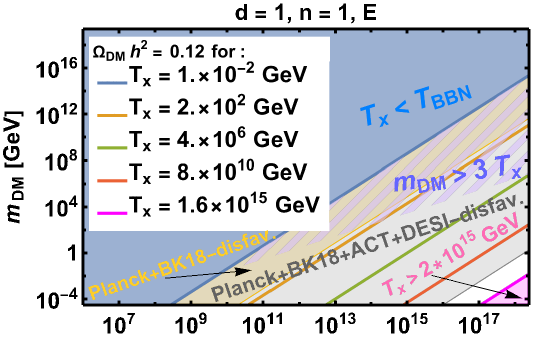} 
\includegraphics[height=5.0cm, trim=14 16 0 0,clip]{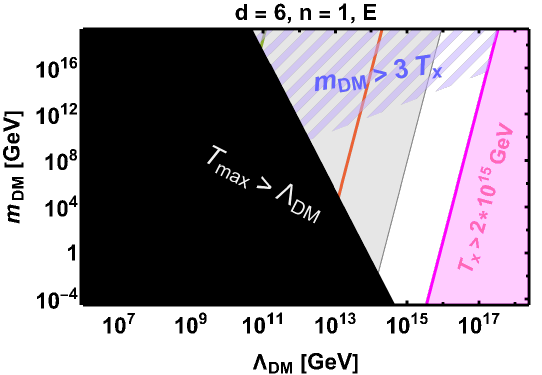} \\[6pt]
\includegraphics[height=5.5cm]{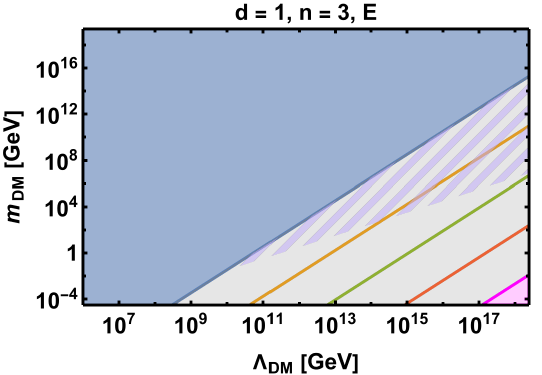} 
\includegraphics[height=5.5cm, trim=14 0 0 0,clip]{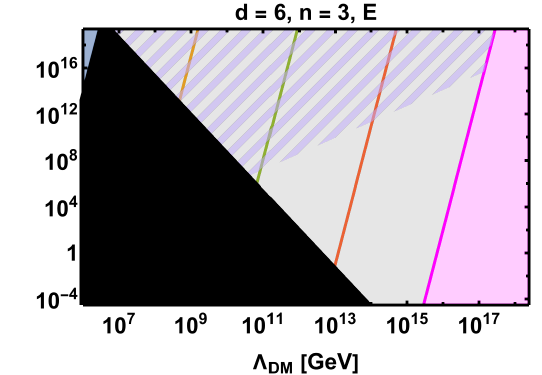} \\[6pt]
\caption{
Various constraints in the parameter space $(\Lambda_{\rm DM}, m_{\rm DM})$ for the prototype model, eq.~\eqref{eq:crosssection}, with E-model, $d=1, 6$, $n=1, 3$. Solid color lines correspond to the values of the reheating temperature necessary for the correct relic abundance, eqs.~\eqref{eq:step}, \eqref{eq:yield}; blue shaded region is excluded by $\Tx < 10^{-2}$\,GeV;   dashed purple shaded region is excluded by $m_{\rm DM}>3 \Tx$; black shaded region is excluded by $T_{\rm max}>\Lambda_{\rm DM}$; 
    yellow (gray) band is disfavored by current Planck+BK18 (Planck+BK18+ACT) data; pink-shaded region is excluded by $\Tx >2\cdot10^{15}$\,GeV.}
    \label{fig:E_model_d_1_d_2_temps_exps}
\end{figure}

There are several comments in order here. Firstly, the period of the universe evolution when the dominant contribution to the production of DM
comes from depends on the dimension of the operator. If it dominantly comes from the reheating period it is sensitive to the maximal temperature $T_{\rm max}$  reached then, before the radiation dominance era (see Figure~\ref{fig:tmaxvstrhplot}). Secondly, such ``theory'' of production makes sense only for $\Lambda_{\rm DM} \gg T_{\rm max}, \Tx$.  And also, and importantly for the presentation of our results,  
we have to remember that we have taken only a single higher dimension operator to describe the DM production, with the numerical coefficient in the above equation taken to be 1. In a concrete model there might be several operators with their Wilson coefficients depending on some powers of the coupling constants, phase space factors etc.. 
Furthermore, in the calculations regarding DM we have assumed instant and local thermal equilibrium to have been reached during the reheating period. This simplification may not be appropriate for some specific models. In such cases, the thermalization process should be analyzed in more detail, as shown in refs.~\cite{Harigaya:2014waa,Harigaya:2019tzu}.
Thus, our results serve merely as an illustration of a general link between CMB data, reheating temperature and DM production mechanism.

With regard to the dependence of the DM abundance on $T_{\rm max}$ during the reheating
period, the question has been investigated in detail in refs.~\cite{Chung:1998rq,Giudice:2000ex,Bernal:2019mhf}.
We  show the results for the production amplitudes of dim $4+d=5$ and $4+d=10$.

In the first case, the production during the reheating period can approximately be neglected, in the second one it is the dominant one \cite{Bernal:2019mhf,Bernal:2019lpc}. Correspondingly, in the first case the DM abundance depends on $\Tx$. In the second case it depends on $T_{\rm max}$ but can still be parametrized  by $\Tx$.  
The number density of DM particles produced via scatterings of particles in the thermal bath, with $\langle\sigma v\rangle$  of the processes given by \eqref{eq:crosssection},  is
determined by the Boltzmann equation
 \begin{equation}
 \label{eq:DMBE0}
	\frac{\d n_\text{DM}}{\d t} + 3H\,n_\text{DM}  = \langle\sigma v\rangle\big((n_{\rm DM}^{\rm eq})^2-n_\text{DM}^2\big)\,.
\end{equation}
Here, $n_{\rm DM}^{\rm eq}$ denotes the equilibrium DM number density.
In this paper we assume that this is the only process for production of the DM and that initially $n_{\rm DM}=0$. In particular, we assume that the inflaton direct decays to DM can be neglected. In a more general analyses those assumptions can be relaxed.\footnote{
Thermalization and number-changing processes within the dark sector may modify the DM abundance ~\cite{Chu:2013jja, Bernal:2015ova, Bernal:2015xba, Bernal:2017mqb, Falkowski:2017uya, Heeba:2018wtf, Mondino:2020lsc, Bernal:2020gzm}, we do not consider such cases as well.}

\begin{figure}[tbp]
   \includegraphics[height=5.cm]{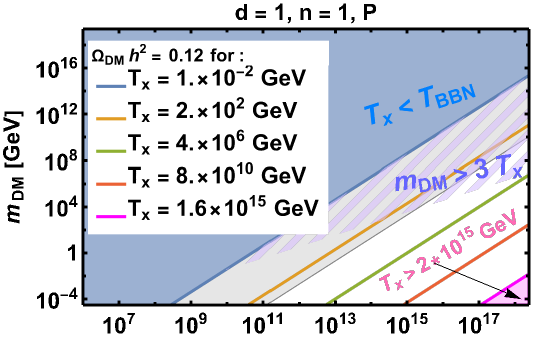} 
\includegraphics[height=5.cm, trim=14 0 0 0,clip]{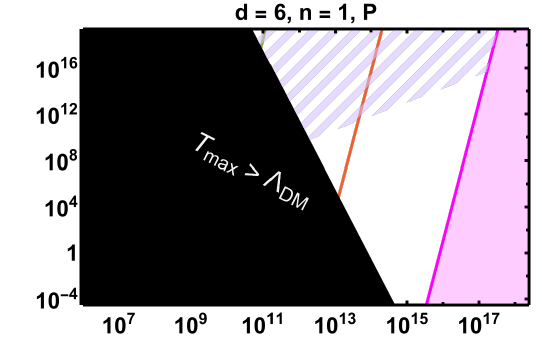} \\[6pt]
\includegraphics[height=5.55cm]{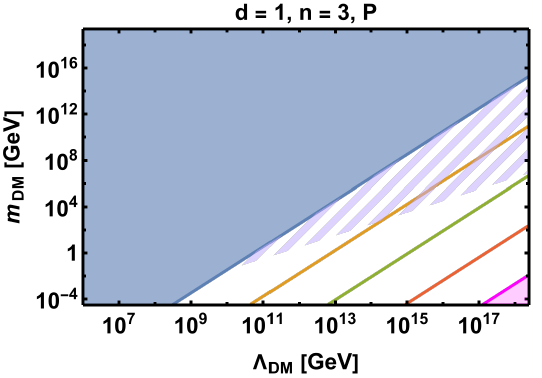} 
\includegraphics[height=5.55cm, trim=14 0 0 0,clip]{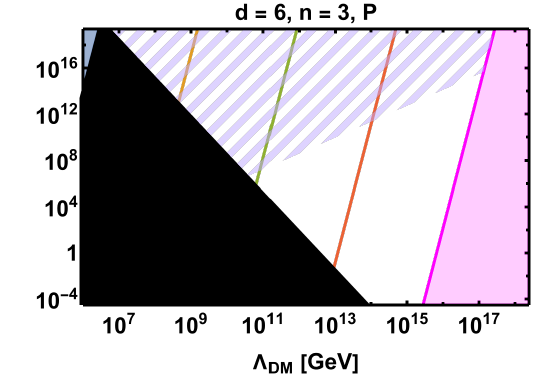} 
    \caption{
 Plots analogous to those in Figure~\ref{fig:E_model_d_1_d_2_temps_exps} for P-model.}
    \label{fig:P_model_d_1_6_temps}
\end{figure}

For the comoving DM particle density  $Y\equiv n_\text{DM}/s$ one has
\begin{equation}
\label{eq:DMBE}
	\frac{\d Y}{\d T} =\frac{\langle\sigma v\rangle\,s}{H\,T}\left(Y^2-Y_\text{eq}^2\right)\,,
\end{equation}
where $s$ is the entropy density 
\begin{equation}
	s(T)\equiv \frac{2\pi^2}{45}\,g_{*s}(T)\,T^3
\end{equation}
with $g_{*s}(T)$ being the effective number of relativistic degrees of freedom in the SM plasma~\cite{Drees:2015exa} 
The DM relic abundance has been calculated
by numerical integration. 
However,
for a qualitative understanding of the results it is useful to present
its approximate analytical solutions. At a very qualitative level, since DM never reaches the equilibrium distribution in freeze-in production, i.e.~$Y\ll Y_\text{eq}$, we have ~\cite{Elahi:2014fsa}
\begin{equation}
Y\sim \int^{\widetilde{T}}_0\frac{\Mp\,T^{2(d-1)}}{\Lambda_{\rm DM}^{2d}}\sim\frac{\Mp\,\widetilde{T}^{2d-1}}{\Lambda_{\rm DM}^{2d}}~.
\label{eq:step}
\end{equation}
The upper limit in the integral, $\widetilde{T}$, is in principle given
by the maximal temperature $T_{\rm max}$ reached during the reheating process. However, as we already above discussed, the relative importance of the contribution to the DM abundance coming from the reheating period respectively to the post-heating time, depends on the dimension of the operator describing the production process. For low dimension operators, for instance $d=1,2$, the reheating contribution is negligible and to a very good approximation the upper limit of integration can be taken as $\widetilde{T}=\Tx$. For bigger values of $d$ we use $\widetilde{T}=T_{\rm max}$.

The DM yield $Y$ is related to the relic density via
\begin{equation}
\Omega_{\rm DM}
=\frac{m_{\rm DM} Y  s_0}{\rho_c}
\simeq
0.2\times \left(\frac{m_{\rm DM}}{1~{\rm TeV}}\right) \left(\frac{Y}{10^{-13}}\right)~,
\label{eq:yield}
\end{equation}
where $\rho_c$ represents the critical density of the Universe, $s_0$ denotes entropy density at present epoch at $T_0=2.75~{\rm K}\sim10^{-4}~{\rm eV}$. We have also approximated $\sqrt{g_*}g_{*s}\sim10^3$. The Planck experiment gives $\Omega_{\rm DM} h^2\sim 0.12$.

The DM abundance is dependent on the reheating temperature as evident from eq.~\eqref{eq:step}. This feature gives us the scope of correlating the DM phenomenology with the cosmological quantities $(n_s,r)$ as earlier stated, for a chosen model of inflation. Those results  are illustrated in Figures~\ref{fig:E_model_d_1_d_2_temps_exps}-\ref{fig:P_model_d_1_6_temps}
for the ``prototype'' models given by eq.~\eqref{eq:crosssection}.

We show the results as the plots of $m_{\rm DM}$ vs $\Lambda_{\rm DM}$. 
They illustrate two kinds of constraints on our prototype models. Firstly, those are the constraints
following from the DM production mechanisms,
and secondly, those following from the requirement of consistency with the chosen inflationary model  and the CMB observables.

To the first category belong colored thin solid lines representing the $\Tx$ values required to satisfy the observed relic density of DM. Along with this, there are denoted regions excluded by theoretical consistency conditions, namely for $m_{\rm DM} > 3 \Tx$ needed for the assumption of the produced DM particles to be relativistic and $T_{\rm max}<\Lambda_{\rm DM}$ for the validity of EFT of UV-freeze-in DM. Moreover, there are also shown the model independent bounds on the reheat temperature of the universe $\Tx$, as in eq.~\eqref{eq:Tx-bounds}, translated into the ($\Lambda_{\rm DM}$, $m_{\rm DM}$) parameter space.
To the second category belong
the constraints on the reheating temperature obtained in the considered  inflationary models confronted in Section~\ref{sec:CMB} with the CMB observables from Planck and Planck-ACT observations. 

In all cases shown in Figures~\ref{fig:E_model_d_1_d_2_temps_exps}-\ref{fig:P_model_d_1_6_temps} a large region of the parameter space is ruled out from the requirement of theoretical consistencies and by the model independent bounds on the reheating temperature. The main effect visible here is the dependence of the constraints on the dimension $d$ of the operators responsible for DM production process: the higher the dimension $d$ the stronger the dependence on the reheating temperature (eq.~\eqref{eq:crosssection}). 
In consequence, value of $m_{\rm DM}$ necessary to get the correct relic abundance for a given $T_{\rm max}$ (so also given reheat temperature $\Tx$ -- see Figure~\ref{fig:tmaxvstrhplot}) grows faster with $\Lambda_{\rm DM}$ when $d$ is bigger. In addition, stronger dependence on $T_{\rm max}$ (and $\Tx$) results in bigger distances between lines of two fixed temperatures. As a result, the regions excluded by the condition $T_{\rm max}<\Lambda_{\rm DM}$
(black regions in Figures~\ref{fig:E_model_d_1_d_2_temps_exps}-\ref{fig:P_model_d_1_6_temps}) are bigger for bigger $d$.
For big values of $d$ there is additional dependence on the value of the exponent $n$. It follows from the fact that the inflaton energy density red-shifts faster when $n$ is bigger (at early stages of reheating $\rho_\phi\propto a^{-3(1+w)}$ with $w$ given by eq.~\eqref{w}). Faster decrease of $\rho_\phi$ gives faster decrease of $H$ which results in bigger coefficient in eq.~\eqref{eq:DMBE}. Thus, for bigger $n$ we obtain bigger DM yield so smaller $m_{\rm DM}$ is needed to obtain the observed relic abundance of DM.

In the remaining regions the second category of constraints  becomes relevant and the picture is more diversified. The   constraints  on the reheating temperature from the CMB observables, from either Planck+BK18 data or from Planck+BK18+ACT+DESI data give some dependence on the inflationary models and particularly on the value of $n$. The details of the plots can be understood with help of Figures~\ref{fig:nsrPlotEmodel} and \ref{fig:nsrPlotTPmodels}. 
 We stress again that the results presented here are for the prototype model, eq.~\eqref{eq:crosssection}, and for a concrete complete model such an analysis has to be redone.   
For instance a numerical factor in the r.h.s.~of eq.~\eqref{eq:crosssection} different from 1 corresponds to rescaling $\Lambda_{\rm DM}$ by the square root of such factor, which would result in an appropriate shift on the horizontal axes in Figures \ref{fig:E_model_d_1_d_2_temps_exps} and \ref{fig:P_model_d_1_6_temps}. However, our results illustrate potential usefulness of such an analysis for  concrete models.

Many freeze-in models with the DM production described by dim 5 and dim 6 effective operators have been discussed in the literature. They include, for instance, two scalar thermal bath particles $\Phi$ annihilating into a pair of DM fermions $\chi$, $\mathcal{L}=|\Phi|^2\bar{\chi}\chi /\Lambda_{\rm DM} $, or two thermal bath fermions $\Psi$ annihilating either into a pair of scalar DM particles $S$, ${\cal L}={\bar \Psi}\Psi SS /\Lambda_{\rm DM}$ or into a pair of DM fermions $\chi$, ${\cal L}={\bar \Psi}\Psi{\bar\chi}\chi/\Lambda_{\rm DM}^2$.

This has been studied in the context of Higgs-portal, Spin 2 -portal, gravitino-portal, sterile neutrino-portal, scotogenic DM  where DM loop exchange generates tiny SM neutrino mass, with correlation with neutrino oscillation data, axion and axino-portal, see ref.~\cite{Bernal:2017kxu,Bernal:2019mhf} for a review. One important scenario in our context is the gravitational production scenarios, with dim 5 or dim 6 effective operators \cite{Koutroulis:2023fgp,Lebedev:2022cic}, 
where simply $\Lambda_{\rm DM}=\Mp$ in Figures~\ref{fig:E_model_d_1_d_2_temps_exps}-\ref{fig:P_model_d_1_6_temps}. 
These models have their characteristic dependencies on the reheat temperature $\Tx$ and our prescription presented in the paper will readily probe those scenarios.

\medskip
\section{Conclusions}

The present and future experimental programmes will explore the properties of the Cosmic Microwave Background in a great detail.
In this paper we have proposed a systematic
approach to use the CMB observables and the model-independent bounds on the reheating temperature  a) for discriminating between different models of inflation, b) to constrain a reheating temperature dependent particle physics scenario, freeze-in DM production models. In the first step of this approach, all  independent parameters of an inflationary model are expressed in terms of the CMB observables. Next, adopting a one parameter effective description (by an effective dissipation rate $\Gamma$), a consistency relation fixes the reheating temperature in terms of the measured values of the CMB observables. For a given model of inflation the bounds on those observables
then follow from model-independent bounds on the reheating temperature (see Section~\ref{sec:CMB}). The striking consequence of those bounds is that a given model of inflation gives very narrow ranges of the  power spectrum index $n_s$, different for different inflationary models and with very interesting dependence
on the value of $r$. 
For some inflaton potentials those ranges are further narrowed when effects of non-perturbative fragmentation of the inflaton condensate are taken into account.
In this paper, the above described strategy has been applied to $\alpha$-attractor inflaton potentials, 
with three free parameters.
Most of the considered models give predictions consistent with the Planck+BK18 data for wide ranges of parameters. Inclusion of ACT and DESI data leads to somewhat bigger values of the spectral index $n_s$ so prefers some class of models.
The biggest value of $n_s$ may be obtained in the polynomial $\alpha$-attractor P-model with $n=1$ and high reheat temperature, corresponding to $N_k\approx55$.\footnote{We thank Renata Kallosh and Andrei Linde for the dicussion on this point.}
We stress that those results are obtained for a simple reheating model
described by a constant decay rate of inflaton and with non-perturbative fragmentation of the inflaton condensate taken into account.

Next, expressing the reheating temperature $\Tx$ in terms of the CMB observables gives an interesting probe of the $\Tx$ dependent post-inflationary processes by the CMB observables. In this context  UV-freeze-in  DM production models have been discussed, with the DM production process described
by higher dimensional  effective operators. The impact of the measured CMB observables may give interesting constraints on the DM candidate mass versus the scale of the effective operators and on the consistency 
of the DM production models with models of inflation. However, we also observe that important role in constraining the ($\Lambda_{\rm DM}$, $m_{\rm DM}$) parameter space is played by such theoretical constraints like the validity of EFT and 
the assumed relativistic nature of DM.  We observe that for concrete values of the scale $\Lambda_{\rm DM}$ there are upper bounds on $m_{\rm DM}$, strongly dependent on the dimension of the  EFT operator and on the parameter $n$ of the inflaton potential (for instance in our ``prototype'' models for the dimension five operators, they never exceed ${\cal{O}}(10^{7})$\,GeV, and are much lower for low values of $\Lambda_{\rm DM}$ (see Figure~\ref{fig:P_model_d_1_6_temps}). 

Our approach can be extended in several interesting directions, such as for example
a) inclusion of a hidden sector, with its own reheating temperature, b) more complex reheating mechanisms c) more parameters in the inflaton potential d) other $\Tx$ sensitive  processes, for instance freeze-out during the reheating period. 

Our final remark is that with the advent of Gravitational Wave (GW) astronomy, beyond the CMB experiments, to pulsar timing array and interferometer based GW detectors, we may aspire to achieve complementary probes of inflationary phenomenology and the reheating era. The first step towards this direction in the context of DM and reheating has been taken 
in refs.~\cite{Ghoshal:2024gai,Cheek:2025gvx}.

\section*{Acknowledgments}
The authors would like to thank Renata Kallosh and Andrei Linde for many discussions. 
A.G.~thanks S K Jeesun for comments.
P.K.~thanks Adeela Afzal and Shiladitya Porey for sharing their Mathematica codes used to produce the experimental contours in Section~\ref{sec:CMB}.

\appendix

\section{Formulae for \boldmath $\alpha$-attractor T- and P-models}
\label{app:TP-models}

Formulae related inflaton potential parameters with the CMB-related quantities for $\alpha$-attractor E-model were presented in Section~\ref{sec:CMB}. Here we present analogous formulae for other $\alpha$-attractor models.

From the potential of the T-model
\begin{equation}
V_T(\phi) = \Lambda_{\rm inf}^4 
\left(\tanh\left(\sqrt{\frac{2}{3\alpha}}\,\frac{\phi}{\Mp}\right)\right)^{2n}
\end{equation}
the following expressions for the CMB parameters follow
\begin{eqnarray}
A_s
&=&
\frac{\alpha}{128\pi^2n^2}\frac{\Lambda_{\rm inf}^4}{\Mp^4}
\left(\sinh\left(2\sqrt{\frac{2}{3\alpha}}\,\frac{\phi_k}{\Mp}\right)\right)^{2}
\left(\tanh\left(\sqrt{\frac{2}{3\alpha}}\,\frac{\phi_k}{\Mp}\right)\right)^{2n}
\,,\\
n_s
&=&
1-\frac{32n}{3\alpha}\,\left(n+\cosh\left(2\sqrt{\frac{2}{3\alpha}}\,\frac{\phi_k}{\Mp}\right)\right)\left(\sinh\left(2\sqrt{\frac{2}{3\alpha}}\,\frac{\phi_k}{\Mp}\right)\right)^{-2}
\,,\\
r
&=&
\frac{256 n^2}{3 \alpha}  
\left(\sinh\left(2\sqrt{\frac{2}{3\alpha}}\,\frac{\phi_k}{\Mp}\right)\right)^{-2}
\end{eqnarray}
They may be inverted giving the results
\begin{eqnarray}
\alpha
&=&
\frac{256n^2}{3r}\frac{1}{\xi^2-1}
\,,\\
\phi_k
&=&
\Mp\sqrt{\frac{32}{r\left(\xi^2-1\right)}}\,n\,\ln\left(\xi+\sqrt{\xi^2-1}\right)
\,,\\
\Lambda_{\rm inf}^4
&=&
\frac{3\pi^2}{2}\,\Mp^4\,rA_s\left(\frac{\xi+\sqrt{\xi^2-1}+1}{\xi+\sqrt{\xi^2-1}-1}\right)^{2n}
\,,\\
\alpha_s
&=&
-\frac{r^2\left(2\xi^2+2\xi\sqrt{\xi^2-1}-1\right)\left(2n\xi+\xi^2+1\right)\left(\xi+\sqrt{\xi^2-1}\right)^4}{8n^2\left(\xi\sqrt{\xi^2-1}\left(16\xi^4-16\xi^2+3\right)+\left(2\xi^2+1\right)\left(16\xi^4-16\xi^2+1\right)\right)}\,.
\end{eqnarray}
The condition $\epsilon=1$ (``usual'' end of inflation) leads to the following values of the inflaton field $\phi_{\rm end}$, the energy density $\rho_{\rm end}$ and the number of e-folds during inflation after the pivot scale crossed the horizon $N_k$: 
\begin{eqnarray}
\phi_{\rm end}
&=&
\Mp\,\frac{4\sqrt{2}\,n}{\sqrt{r(\xi^2-1)}}\,\ln\left(\frac14\left(\sqrt{r(\xi^2-1)}+\sqrt{16+r(\xi^2-1)}\right)\right)
,\\
\rho_{\rm end}
&=&
2\pi^2\Mp^4\,r\,A_s\,
\left(\frac{\left(\xi+\sqrt{\xi^2-1}+1\right)\left(\sqrt{r(\xi^2-1)}+\sqrt{16+r(\xi^2-1)}-4\right)}{\left(\xi+\sqrt{\xi^2-1}-1\right)\left(\sqrt{r(\xi^2-1)}+\sqrt{16+r(\xi^2-1)}+4\right)}\right)^{2n},
\\
N_k
&=&
\frac{8n}{r(\xi^2-1)}
\left(2\xi
-\frac{\sqrt{16+r(\xi^2-1)}+\sqrt{r(\xi^2-1)}}{4}
-\frac{4}{\sqrt{16+r(\xi^2-1)}+\sqrt{r(\xi^2-1)}}
\right).
\nonumber\\
\end{eqnarray}

For P-models with the potential
\begin{equation}
V_P(\phi) = \Lambda_{\rm inf}^4 \frac{\left(\sqrt{\frac{2}{3\alpha}}\,\frac{\phi}{\Mp}\right)^{2n}}
{1+\left(\sqrt{\frac{2}{3\alpha}}\,\frac{\phi}{\Mp}\right)^{2n}}
\end{equation}
one obtains
\begin{eqnarray}
A_s
&=&
\frac{1}{48\pi^2n^2}\frac{\Lambda_{\rm inf}^4\phi_k^2}{\Mp^6}
\left(\frac{2\phi_k^2}{3\alpha \Mp^2}\right)^{n}
\left(1+\left(\frac{2\phi_k^2}{3\alpha \Mp^2}\right)^{n}\right)
\,,\\
n_s
&=&
1-4n\frac{\Mp^2}{\phi_k^2}\,
\left(n+1+(2n+1)\left(\frac{2\phi_k^2}{3\alpha \Mp^2}\right)^{n}\right)
\left(1+\left(\frac{2\phi_k^2}{3\alpha \Mp^2}\right)^{n}\right)^{-2}
\,,\\
r
&=&
32n^2\frac{\Mp^2}{\phi_k^2}\,
\left(1+\left(\frac{2\phi_k^2}{3\alpha \Mp^2}\right)^{n}\right)^{-2}
\end{eqnarray}
and
\begin{eqnarray}
\alpha
&=&
\frac{64n^2}{3r}\left(\frac{2n+1}{2n+\xi}\right)^2\sqrt[n]{\frac{2n+1}{\xi-1}}
\,,\\
\phi_k
&=&
\Mp\sqrt{\frac{32}{r}}\,n\,\frac{2n+1}{2n+\xi}
\,,\\
\Lambda_{\rm inf}^4
&=&
\frac{3\pi^2}{2}\,\Mp^4\,rA_s\left(\frac{\xi-1}{2n+\xi}\right)
\,\\
\alpha_s
&=&
-\frac{r^2\left(\xi^2(n+1)+2n(\xi+n)\right)}{64n^2(2n+1)}
\,.
\label{eq:alphaS-P}
\end{eqnarray}
It is not possible to get closed expression for $\phi_{\rm end}^{(\epsilon)}$ in P-models because it is related to the solution of equation $x(1+x^{2n})={\rm const}$ (explicit solutions in the case of $n=1$ can be found but they are rather lengthy and complicated).

\medskip

\section{Approximations and simplifying assumptions}
\label{app:approx}

In this appendix, we discuss the approximations we used and evaluate the accuracy of the results obtained using them.

Three approximations were used in Section \ref{sec:CMB} to obtain analytical expressions describing the reheating process:
\begin{itemize}
\item
Inflaton potential during the reheating was approximated by the appropriate monomial of the infalton field i.e.~the first term of the Taylor expansion of the full potential \eqref{eq:alpha-attractors}.
\item
Contribution of the radiation energy density to the Hubble parameter during the reheating was neglected (see eq.~\eqref{H_approx}).
\item
Contribution of the inflaton energy density after the reheating process was neglected.
\end{itemize} 
All the above approximations introduce some modifications of energy densities and number of e-folds for considered processes. This influences the results obtained for the reheating temperature $T_\times$ and the total number of e-folds necessary to apply the consistency condition \eqref{eq:k=aH}. Which in turn has some impact on the relations between $T_\times$ and the CMB parameters $n_s$ and $r$ shown in Figures \ref{fig:ns-r_Tx-Nk}-\ref{fig:nsrPlotTPmodels}. In order to estimate the accuracy of our results we compared them with the results of numerical solutions of appropriate Boltzmann equations for several cases. For each type of model (E, T and P) we compared analytical and numerical results for the spectral index $n_s$ for several combinations of values of $n$, $r$ and $T_\times$. In all cases values of $n_s$ obtained with numerical integration are slightly smaller than those resulting from our analytical expressions. However, the differences are very small. In the case of E- and T-models we obtained $\Delta n_s$ in the range of about $(2\div3)\cdot10^{-4}$. For P-models the correction to $n_s$ is similar for most cases and is bigger only for relatively big values of $r$ and big $n$. For $r\approx0.01$ the correction $\Delta n_s$ grows to about $5\cdot10^{-4}$ for $n=5$ and is almost $1\cdot10^{-3}$ for $n=10$. But, as visible in the right lower panel in Figure \ref{fig:nsrPlotTPmodels}, P-model with $n=10$ and $r$ of order $0.01$ predicts quite small values of $n_s$ not favored by present experiments. Moreover, such model with relatively big values of $r$ is characterized by rapid variability of $n_s$, so correction of order $10^{-3}$ would be hardly visible. In summary, using numerical results would systematically shift all the curves of fixed $T_\times$ shown in Figures \ref{fig:ns-r_Tx-Nk}-\ref{fig:nsrPlotTPmodels} by about 0.00025 (with exact value depending slightly on the parameters) to the left.

We have defined the end of the reheating and beginning of the RD as the moment $t_\times$ at which the energy density of radiation equals one half of the total energy density. It is in full analogy to the standard terminology in which the end of RD and the beginning of MD is defined by the moment $t_{\rm eq}$ at which energy density of matter equals one half of the total energy density.
Quite often in the literature another definition is used, according to which reheating ends at $t_{H=\Gamma}$ when $H=\Gamma$. Such definition simplifies some calculations but it leads to the result that the contribution of radiation to the total energy density depends on the shape of the inflaton potential. Using our analytical formulae from Section \ref{sec:CMB} we find that in such case $\rho_{\rm R}(t_{H=\Gamma})/\rho_{\rm tot}(t_{H=\Gamma})\approx9/(29-12w)$. However, the difference between temperatures of radiation at $t_\times$ and $t_{H=\Gamma}$ are quite small, namely
\begin{equation}
T_{H=\Gamma}\approx\frac{\sqrt{3}\,(5-3w)^{1/4}}{(2+\sqrt{9-3w})^{1/2}}\,T_\times\,.
\label{eq:TxTGamma=H}
\end{equation}
The ratio of $T_{H=\Gamma}/T_\times$ decreases with $n$ from about 1.16 for $n=1$ to only about 1.03 for $n=10$. So small modification of (definition of) the reheating temperature would result in the change of the corresponding value of $n_s$ at most of the order $10^{-5}$.

In the Section \ref{sec:DM} we have also made several simplifying assumptions:
\begin{itemize}

\item 
We have assumed that the tree level and loop level decays of the inflaton to DM particles are negligible.

\item 
We have assumed the gravitational production of DM during inflation to be negligible, see ref.~\cite{Ghoshal:2024gai} for a discussion when it is justified.

\item 
In the calculations regarding DM we have assumed instant and local thermal equilibrium to have been reached during the reheating period. 
\end{itemize}

\medskip

\bibliographystyle{BiblioStyle}

\bibliography{CMB-Tx-DM_JCAP_ref}

\end{document}